\documentclass[%
 reprint,
 amsmath,amssymb,
 aps,
 pra,
]{revtex4-2}

\usepackage{graphicx}
\usepackage{dcolumn}
\usepackage{bm}
\usepackage{subfig}

\begin{document}

\preprint{APS/123-QED}







\title{The influence of phonon harmonicity on spectrally pure  resonant Stokes fields}


\author{Georgios Stoikos}
 \altaffiliation[Also at ]{School of Applied Mathematics and Physical Sciences, National Technical University of Athens.}
\author{Eduardo Granados}%
 \email{eduardo.granados@cern.ch}
\affiliation{%
CERN, 1217 Geneva, Switzerland
}


\date{\today}

\begin{abstract}

Thanks to their highly coherent emission and compact form factor, single axial mode diamond Raman lasers have been identified as a valuable asset for applications including integrated quantum technology, high resolution  spectroscopy or coherent optical communications. While the fundamental emission linewidth of these lasers can be Fourier limited, their thermo-optic characteristics lead to drifts in their carrier frequency, posing important challenges for applications requiring ultra-stable emission. We propose here a method for measuring accurately the temperature-dependent index of refraction of diamond by employing standing Stokes waves produced in a monolithic Fabry-P\'{e}rot (FP) diamond Raman resonator. Our approach takes into account the influence of the temperature on the first-order phonon line and the average lattice phonon frequency under intense stimulated Raman scattering (SRS) conditions.  We further utilize this model to calculate the temperature-dependent thermo-optic coefficient and the Gr{\"u}neisen parameter of diamond in the visible spectral range. The theory is accompanied by the demonstration of tunable Fourier-limited Stokes nanosecond pulses with a stabilized center frequency deviation of less than $<$4 MHz.




\end{abstract}

\maketitle


\section{\label{sec:level1}Introduction\protect}

Diamond has been fascinating humanity for centuries, both for its rarity and its exceptional physical properties. Due to modern breakthroughs in synthetic diamond production, mostly thanks to chemical vapour deposition (CVD) technology, diamond is now widely available for a variety of scientific applications \cite{doi:https://doi.org/10.1002/9783527648603.ch1}.  The everlasting interest in diamond continues today in step with the development of new applications across different fields of science. As such, diamond holds promise to be one of the premier materials for quantum applications \cite{TERAJI202037}, including quantum computing, generation of single photons \cite{NEU2014127}, quantum sensing \cite{Markham:19} and quantum memories \cite{PhysRevLett.119.223602}.


Knowledge of diamond's optical and mechanical properties runs deep for most factors with an exception being the temperature dependence of the refractive index (also known as thermo-optic coefficient), despite its importance for optical applications or integrated photonic devices in diamond. In the literature it is usually found as a single value of $(1/n)\partial n/\partial T=5\times 10^{-6}$~K\textsuperscript{-1} at 300~K for the low-frequency limit \cite{Fontanella:77}, or in the far-infrared range for a temperature range of up to 925~K \cite{PhysRevB.62.16578}. Information regarding the thermo-optic coefficient of diamond at visible wavelengths and at extended temperature ranges remains elusive.

A thorough theoretical description on the thermo-optic coefficient in diamond is a great challenge since it requires a working model for the dielectric function and its re-normalization by the electron-phonon interaction and the thermal expansion of the lattice \cite{PhysRevB.62.16578}. There have been models using empirical pseudo-potentials for the thermo-optic coefficients of different semiconductors but since C has a large Debye temperature  $\Theta_D=1880$~K, knowledge of the thermo-optic coefficient at $300$~K does not provide meaningful information for higher temperatures \cite{PhysRevB.2.3193}. The inaccuracy is caused by the fact that a linear approximation of the temperature-dependent index of refraction $n(T)$ is permitted at temperatures much larger than $\Theta_D$ where the material can be described by a single oscillator frequency. 

Knowledge of the thermo-optic coefficient is of paramount importance for applications and the present research was conducted with the purpose of gaining insight in the accurate prediction of the tuning the Stokes resonant frequency in integrated monolithic diamond resonators. This was done with the purpose of employing diamond resonators in quantum technology and spectroscopy. Unfortunately, the existing approximations for the index of refraction render them insufficient for the level of accuracy required  in these applications.

When it comes to the refractive index of diamond, there have been many works relative to the optical and Raman properties of diamond, but there is a scarcity of information regarding the index of refraction under strong vibrational fields and at different temperatures. Ruf et al.'s work provided valuable information in \cite{PhysRevB.62.16578} for estimating the thermo-optic coefficient, however it did not take into account the overall contribution of the different temperature-dependent vibrational modes that can be produced in diamond and thus it needs to be expanded.

We propose here an alternative methodology for measuring and calculating the thermo-optic coefficient. Our methodology includes the use of a monolithic single-frequency FP diamond Raman resonator operating at visible wavelengths. Here, the accurate measurement of the output Stokes frequency as a function of temperature allowed us to retrieve the temperature-dependent index of refraction that produced a frequency shift in the Stokes field output. For this method to be efficient, the use of diamond Raman monolithic resonators is important because it provides an ultra-stable environment that depends exclusively on the resonator opto-mechanical properties and not the pumping laser characteristics or the environment. 

Stimulated Raman Scattering (SRS) provides key advantages when compared to other resonant measurement methods: first, it is not subjected to spatial hole burning and therefore it produces an intrinsically stable single-frequency output \cite{Lux:16}, and second, it is an automatically phase-matched nonlinear process. This means that there is no direct relation between the phases of pump and Stokes waves, and therefore many characteristics of the Stokes output are correlated with the material properties rather than the pump laser allowing us to study the material in a decorrelated manner.



\section{Measurement principle}

In terms of using the diamond bulk as a Raman laser material, its unique optical properties enabled the development of lasers operating over a wide spectrum due to its giant Raman frequency shift (1332~cm\textsuperscript{-1}), large Raman gain ($>$40~cm/GW @ 532~nm) and ultra-wide transparency window (from DUV all the way to the THz, except for a lossy window from 2.6 – 6~\textmu m due to multiphonon-induced absorption \cite{Granados:11,Mildren:09, Spence:10,Lubeigt:10,Sabella:14,Echarri:20}). Furthermore, the excellent thermal properties afforded by diamond (unsurpassed thermal conductivity of 1800~W/m/K at 300~K and low thermo-optic coefficient of the order of 10\textsuperscript{-5}~K\textsuperscript{-1}) along with negligible birefringence \cite{10.1117/12.864981} make it an ideal material for high-power Raman lasing with greatly reduced thermal lensing effects at the kW average power level \cite{Antipov:19}.

The generation of single longitudinal mode (SLM) or narrow linewidth light via SRS in diamond remained elusive until relatively late \cite{Lux:16,Lux:16b, Kitzler:17,Sarang:19,Li:20}. Such bulk cavity systems also require precise alignment, elaborated feedback loops and maintenance of optical components for the laser to function robustly. The further integration of SLM Raman lasers in diamond was recently demonstrated in \cite{Granados22}, showing that by embedding the laser resonator in the Raman media, it was possible to produce frequency stable output from a Fabry-P\'{e}rot (FP) diamond resonator without the need of external mechanical feedback loops to control the cavity length. Moreover, these resonators performed complex functions such as "linewidth squeezing" when pumped by few GHz linewidth multi-mode lasers. Such mechanism, supported by phonon-resonant Raman interactions, directly enhanced the available power spectral density (PSD) of broadband nanosecond lasers by several orders of magnitude.

The frequency stabilization of these FP diamond resonators was carried out by adjusting the temperature of the diamond substrate, which simultaneously influenced the index of refraction, size, and Raman shift of the Raman resonator \cite{doi:10.1063/5.0088592}. In this work we shine light into that complex interplay of thermo-optical and Raman effects by study their dependency on temperature, allowing us to construct a  theoretical model capable of predicting with accuracy the resonating Stokes frequency. Our model is accompanied by an experimental demonstration showing excellent agreement with the proposed theory.

\begin{figure}[ht]
    \centering
    \includegraphics[width=8.6cm]{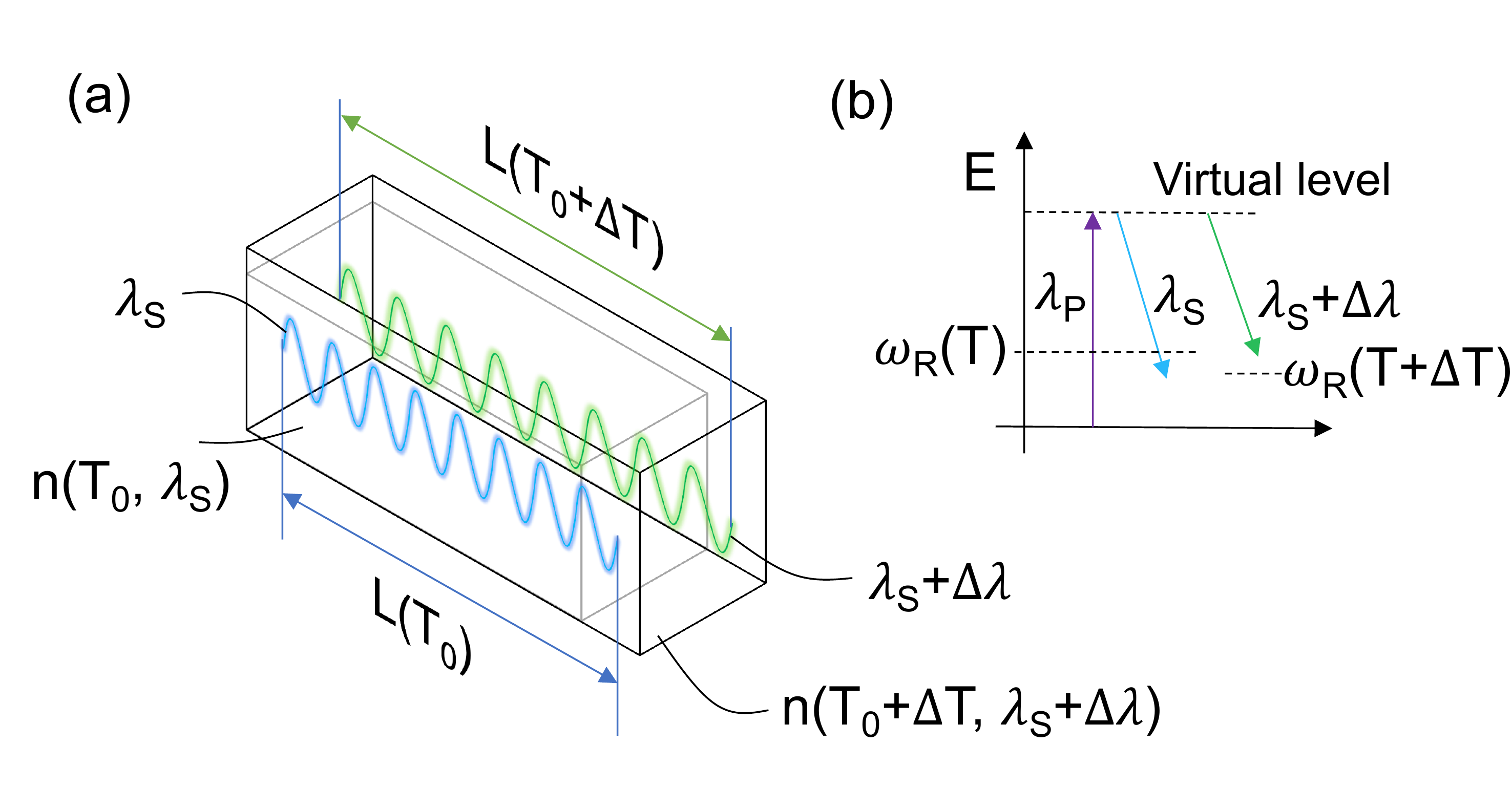}
    \caption{Schematic depiction of: (a) main thermal effects influencing the resonant Stokes frequency in a monolithic Raman resonator, and (b) depicts the temperature effect on the photon and optical phonon energies.}
    \label{fig:thermal_effects}
\end{figure}

We start by identifying the main factors affecting the resonant Stokes frequency as well as their temperature dependency. Those are depicted in Fig. \ref{fig:thermal_effects}, where we have separated the temperature effects on the material optical properties and size (Fig. \ref{fig:thermal_effects}(a)), and on the Raman shift center frequency $\omega_R$ (Fig. \ref{fig:thermal_effects}(b)). In terms of the Stokes resonating frequency, the index of refraction depends simultaneously on the temperature and the chromatic dispersion ($n(T,\lambda)$) due to the shifted Stokes wavelength $\lambda_S' = \lambda_S + \Delta \lambda(T)$. Note that all wavelengths used in this work are in vacuum. The thermal expansion process simultaneously affects the resonating wavelength due to the variable boundary condition (the diamond length $L(T)$ shifts to $L(T+\Delta T)$). Likewise, the Raman shift center frequency $\omega_R(T)$ tuning with temperature does not establish the resonating wavelength nor the tuning slope as a function of temperature but it does affect the location of the boundaries of the longitudinal mode hopping in frequency. We describe all these effects in detail in the following sections, followed by the experimental results.



\section{\label{sec:level1}Single frequency operation of monolithic Fabry-P\'{e}rot diamond Raman resonators \protect}

The production of single frequency resonant Stokes fields depends on many factors, but most importantly on the characteristics of the pump laser intensity, wavelength and linewidth, the resonator optical length, and the temporal envelope of the inter-playing pump and Stokes pulses.

The temporal features of these laser fields -- both in amplitude and phase on timescales shorter than the resonator round-trip time, is caused by the interference of its spectral longitudinal modes. These modes, however, generally vary in amplitude and phase on the timescale of the round-trip time or slower. Since the longitudinal modes vary slowly (much slower than the phonon dephasing time, in diamond 6.8 ps), we can use steady-state Raman theory, even if interference of the modes produces structures that would need transient Raman theory if modelled in the time domain. This approach has been used widely used to analyze SRS with broad-band lasers \cite{SPENCE20171, Dzhotyan_1977, Sidorovich_1978, 1070054, Warner:86, PhysRevA.36.4835, Xiong:07}, and here we employ this method to model the diamond Raman resonator.

To construct the frequency domain model we start by writing the general equations for a fundamental field (or 'pump') with $2m+1$ modes spaced in frequency by $\Omega_F$, and a multi-mode Stokes field with $2m+1$ modes spaced in frequency by $\Omega_F$:

\begin{equation}
\Tilde{E}_F = \sum_{l=-m}^{m}F_l e^{i(\omega_{F(l)}t-k_{F(l)}z)}+cc
\label{eq:1}
\end{equation}

\begin{equation}
\Tilde{E}_S = \sum_{l=-m}^{m}S_l e^{i(\omega_{S(l)}t-k_{S(l)}z)}+cc 
\label{eq:2}
\end{equation}

in which $cc$ represents the complex conjugate of the preceding term,  $\omega_{S(l)} = \omega_{S(0)} + \Omega_F l$, and   $\omega_{F(l)} = \omega_{F(0)} + \Omega_F l$. In these equations, $S_l$ and $F_l$ are complex amplitudes describing the amplitude and phase of the modes travelling inside the diamond. The approximations for the mode wavevector $k_{F(l)} \approx k_{F(0)} + \Omega_F l / u_F$ accounts for the group velocity difference between the fundamental wavepackets, but neglect group velocity dispersion within each wavepacket. And analogously for the Stokes field the mode wavevector $k_{S(l)} \approx k_{S(0)} + \Omega_F l / u_S$. 

In the following we assume that the central Stokes mode $S_0$ is centered within the Raman gain linewidth so that it accesses the highest or monochromatic Raman gain. Note that in our model the following identity is always true:

\begin{equation}
\omega_{F(l)} = \omega_{S(l)} + \omega_R
\label{eq:3}
\end{equation}

here $\omega_R$ is the frequency of the Raman shift at the line center. This, essentially, is to say that the modes of fundamental and Stokes fields are paired. In order to describe the coupling between this set of modes, we rely on steady-state Raman formalism and write it in a non-degenerate mode for four generic modes $F_{l_1}$, $F_{l_2}$, $S_{l_3}$ and $S_{l_4}$:

\begin{equation}
\frac{1}{u_S} \frac{\partial S_{l_4}}{\partial t} \pm \frac{\partial S_{l_4}}{\partial z} \propto F_{l_1}(F^*_{l_2}S_{l_3})
\label{eq:5}
\end{equation}

This was interpreted in \cite{SPENCE20171} as two modes ($F^*_{l_2}S_{l_3}$) driving a phonon field and a third mode $F_{l_1}$ scattering off the phonon field to drive a fourth mode $S_{l_4}$. For fundamental and Stokes pulsed fields with many longitudinal modes or broadband modes, in principle all types of interactions can drive a polarization at the frequency of a generic mode $S_{r_2}$, given that they satisfy the equation:

\begin{equation}
\omega_{S_{r_2}} = \omega_{F_{l_1}} - \omega_{F_{l_2}} + \omega_{S_{r_1}}
\label{eq:5b}
\end{equation}

Now equation (\ref{eq:5}) can be used as model to formulate the amplification of a generic Stokes mode $S_l$ and the depletion of the fundamental modes $F_{l}$ as a function of the other three interacting modes as follows:

\begin{widetext}


\begin{equation}
\frac{1}{u_S} \frac{\partial S_{l}}{\partial t} + \frac{\partial S_{l}}{\partial z} = 
2cn_F\epsilon_0\frac{g_0}{2} \sum_{r} \sum_{j} F_{l-r}(S_{j}F^*_{j-r})\frac{\Delta\omega_R}{\Delta\omega_R -ir\Omega_F} e^{i(l-j) \mu_{\pm} \Omega_F)z}
\label{eq:7}
\end{equation}


\begin{equation}
\frac{1}{u_F} \frac{\partial F_{l}}{\partial t} + \frac{\partial F_{l}}{\partial z} = 
-2cn_F\epsilon_0\frac{g_0}{2\eta} \sum_{r} \sum_{j} S_{l-r}(F_{j}S^*_{j-r})\frac{\Delta\omega_R}{\Delta\omega_R + ir\Omega_F} e^{i(j-l) \mu_{\pm} \Omega_F)z}
\label{eq:8}
\end{equation}

\end{widetext}

The parameter $\mu_{\pm}$ is the group delay difference per meter between the fundamental and Stokes waves. The positive part $\mu_{+}$ accounts for co-propagating waves or forward SRS while the negative $\mu_{-}$ for the backward SRS:

\begin{equation}
\mu_{\pm}  = \frac{1}{u_F} \mp \frac{1}{u_S}
\label{eq:9}
\end{equation}

Equations (\ref{eq:7}) and (\ref{eq:8}) account for all the possible interactions between fundamental and Stokes modes. The resulting spectra for the Stokes field is dependant on the relative magnitudes of resonant and non-resonant terms. The resonant terms ($r$~=~0) have the phonon driving term exactly resonant with the phonon frequency and can access the highest gain, while other non-resonant interactions have a detuning $r\Omega_F$ that reduces gain and causes a phase rotation. For our model, both resonant and non-resonant interactions need to be taken into account since $\Delta\omega_R > \Omega_F$.

Degenerate terms ($j = l$) have no phase mismatch terms ($\Delta k = 0$) even in the presence of dispersion, and because of the degeneracy these terms must always have the correct phase to provide gain. Non-degenerate modes, however, can be neglected in dispersive media where the phase mismatch is $\Delta k \approx (l-j) \Omega_F \mu_{\pm}$, and so these terms will oscillate in and out of phase with the waves they drive. Here we assume that dispersion in diamond is large enough in the UV and visible spectral ranges to neglect non-degenerate mixing modes without loss of accuracy. With this approximation, we can rewrite equations (\ref{eq:7}) and (\ref{eq:8}) forcing $j = l$. Likewise, for the specific case of single longitudinal mode pumping, the equations can be further simplified to:

\begin{equation}
\frac{1}{u_S} \frac{\partial S_{l}}{\partial t} + \frac{\partial S_{l}}{\partial z} = 2cn_F\epsilon_0\frac{g_0}{2}  F_{0}(S_{l}F^*_{0})\frac{\Delta\omega_R}{\Delta\omega_R + il\Omega_F}
\label{eq:12}
\end{equation}

\begin{equation}
\begin{split}
\begin{aligned}
&\frac{1}{u_F} \frac{\partial F_{0}}{\partial t} + \frac{\partial F_{0}}{\partial z} = \\
&=-2cn_F\epsilon_0\frac{g_0}{2\eta} \sum_{r} S_{r}(F_{0}S^*_{r})\frac{\Delta\omega_R}{\Delta\omega_R + ir\Omega_F}
\end{aligned}
\end{split}
\label{eq:13}
\end{equation}



The term $\Delta\omega_R/(\Delta\omega_R -ir\Omega_F)$ reduces the gain of off-resonant terms by a Lorentzian factor $1+(r\Omega_F/\Delta\omega_R)^2$, and therefore the most efficient interaction is always for doubly-degenerate resonant interactions. 



Let's assume now that the combination of pump intensity and resonator losses is adequate so that the amplification is highly preferential for the central mode $S_0$, and is capable of effectively deplete the fundamental field. This configuration will produce the minimal linewidth for a given resonator round-trip loss. This can also occur when the FSR of the diamond resonator is larger than the effective Raman gain linewidth. A way of calculating the resulting Stokes linewidth is by further simplifying equations \ref{eq:12} and \ref{eq:13}, by implying $\{ S_l = 0~\forall~l\neq 0 \}$:

\begin{equation}
\frac{1}{u_S} \frac{\partial S_{0}}{\partial t} + \frac{\partial S_{0}}{\partial z} = 2cn_F\epsilon_0\frac{g_0}{2}  |F_{0}|^2S_{0}
\label{eq:14}
\end{equation}

\begin{equation}
\frac{1}{u_F} \frac{\partial F_{0}}{\partial t} + \frac{\partial F_{0}}{\partial z} = -2cn_F\epsilon_0\frac{g_0}{2\eta}  |S_{0}|^2F_{0}
\label{eq:15}
\end{equation}

We can now use equation (\ref{eq:14}) and (\ref{eq:15}) to model the dynamic interplay between fundamental and Stokes waves in the Raman resonator when the fields are located at the Raman gain line center. Intuitively, it is possible to see in equation (\ref{eq:14}) that the temporal envelope of the Stokes field depends only on the amplitude of the fundamental pumping pulse and not its phase, whereas the resonating Stokes wavelength will depend on the resonator geometry and opto-mechanical characteristics. The resulting Stokes linewidth, therefore, will be directly linked to the temporal envelope and dispersion characteristics of the resonator, being relatively straightforward the generation of nearly Fourier-limited pulses. 

For the case of a temperature tuned Stokes frequency, the mismatch between the Raman gain line center and the resonating Stokes field will produce a reduced gain by a factor  $1+(\Delta\omega_S(T)/\Delta\omega_R)^2$ due to the detuned Stokes mode. Here $\Delta\omega_S(T)$ is the frequency shift produced in the resonator due to temperature. The calculation of this shift is described in the next section.

\section{\label{sec:level1}Relationship between the  Stokes center frequency and the refractive index \protect}

Here we present a methodology for the calculation of the Stokes frequency and its tuning slope for a generic monolithic Raman resonator longitudinal mode and its relation to the refractive index. The condition for resonance within the diamond FP resonator is:

\begin{equation}
\nu_S (T_0) = q\frac{c}{2 L_{\text{eff}}(T_0, \nu_S)}
\label{eq:2}
\end{equation}

where $q$ is the mode number, $c$ is the speed of light in vacuum, and $L_{\text{eff}}(T_0,\nu_S)$ is the effective length of the resonator at the Stokes resonating frequency $\nu_S$ and at temperature $T_0$. $L_{\text{eff}}$ can be calculated as $L_{\text{eff}}(T_0, \nu_S) = L(T_0) n(\lambda_S,T_0)$, being $n(\lambda_S,T_0)$ the wavelength-temperature dependent index of refraction, and $L(T_0)$ the resonator physical length at temperature $T_0$. In \cite{PhysRevB.62.16578}, the temperature dependent part of the refractive index $n_T\left(T\right)$ was separated from the wavelength dependent part $n_{\lambda}\left(\lambda\right)$. The two terms are added together to give the total index of refraction as:

\begin{equation}\label{eq:dispersion}
n\left(\lambda, T\right)=n^0_{\lambda}\left(\lambda\right) + n_T\left(T\right)
\end{equation}

We note that $n^0_{\lambda}\left(\lambda\right)$ refers to the Sellmeier equation at 0~K and $n_T\left(T\right)$ represents the change of index due to temperature at a fixed wavelength. For small shifts in temperature ($\Delta T$) we can use a perturbation theory approach to estimate the resulting wavelength shift of the Stokes by:





\begin{equation}
\begin{split}
\lambda_S(T_0 + \Delta T) = \frac{2}{q} \left(L(T_0) + \frac{\partial L}{\partial T} \Delta T \right) \cdot \\
\cdot \left(n(T_0,\lambda_S) + \frac{\partial n}{\partial T} \Delta T + \frac{\partial n}{\partial \lambda} \Delta \lambda_S \right)
\end{split}
\label{eq:3x}
\end{equation}

Here the term $\partial L / \partial T$ can be expressed in terms of the linear thermal expansion coefficient ($\alpha$ in the following) as $ \partial L / \partial T = \alpha L(T_0)$. The shift in wavelength can be directly calculated by $\Delta \lambda_S = \lambda_S(T_0 + \Delta T) - \lambda_S (T_0)$. The terms $\partial n / \partial T$ and $\partial n / \partial \lambda$ correspond to the thermo-optic coefficient at $T_0$ and the chromatic dispersion at $\lambda_S$, respectively. Here we assume that dispersion terms do not change for small temperature increments $\Delta T$. 

Reorganizing equation (\ref{eq:3x}) and neglecting second order differential terms, we can obtain an approximate tuning slope of the center Stokes wavelength as a function of temperature:

\begin{equation}\label{dfdt}
    \frac{\partial \nu_S}{\partial T}\Big|_{T_0}=-c\frac{\frac{1}{n}\frac{\partial n_T}{\partial T}+\alpha(T)}{\lambda_S\left(1-\frac{\lambda_S}{n}\frac{\partial n_\lambda}{\partial \lambda}\right)}
\end{equation}

where $c$ is the speed of light in vacuum, $n$ the index of refraction of diamond at the Stokes wavelength and at temperature $T_0$ and $\alpha(T)$ the temperature dependent thermal expansion coefficient of CVD diamond. Jacobson et al. modeled $\alpha(T)$ in \cite{JACOBSON2019107469}, having the following form: 

\begin{equation}
\alpha(T)=\sum_{i=1}^n X_i \, E\left(\frac{\Theta_i}{T}\right)
\end{equation}

where $E(x)$ corresponds to the function given by:

\begin{equation}
E(x)=\frac{x^2e^x}{\left(e^x-1\right)^2}
\end{equation}

Experimental values for $X_i$ and $\Theta_i$ can be found in reference \cite{JACOBSON2019107469}, and those are the ones used here for the calculations.



Continuing, $\frac{1}{n}\frac{\partial n_T}{\partial T}$ is the thermo-optic coefficient. Typical values for this coefficient in the literature are approximately $5\times10^{-6}$~K$^{-1}$ \cite{Fontanella:77}. 


The key part here is in the understanding of the temperature dependent term $n_T\left(T\right)$, which requires to identify the effects that influence it. In general, the index of refraction depends on the lattice energy, which here it is assumed proportional to the internal energy of the system \cite{HERVE1994609}. For diamond, it is possible to use a Bose-Einstein distribution to describe the unit cell infra-red active vibration \cite{doi:10.1080/00018736400101051}.
As a consequence, we can refer to the approximation of the temperature dependent index of refraction as $n_T\left(T\right)$ described in \cite{PhysRevB.62.16578}:

\begin{equation}\label{nt}
    n_T\left(T\right)=A \left(\frac{1}{e^{\frac{\hbar\omega_0}{kT}}-1}+\frac{1}{2}\right)
\end{equation}

with the first term being the Bose-Einstein distribution. Ruf et al. estimated the values of $A$ and $\hbar\omega_0$ based on fits of the data to their experimental measurements \cite{PhysRevB.62.16578}. Their results retrieved a value of  $A$~=~0.01902 and an average phonon frequency of $\hbar\omega_0~=~711$~cm$^{-1}$, independent of temperature. Even though the results in that work fit well their experiments, the influence of the thermal expansion of the lattice to the vibrational eigenfrequencies was not taken into account, and neither its contribution to the line shift of the average phonon frequency $\omega_0$.

This effect is well known for other crystals such as silicon but has yet to be studied experimentally for diamond. In this work we follow a similar rationale to the one described in \cite{article}, and we express the temperature-dependent average phonon frequency that can be calculated as:

\begin{equation}\label{phonon_energy}
   \omega(T) = \omega_0 + \Delta^{(1)}(T)  
\end{equation}


here $\omega_0$ is equivalent to the Raman shift at 0~K, and $\Delta^{(1)}(T)$ is the thermal-expansion contribution. Note that we did not consider coupling of phonons to higher order multi-phonon states, which in principle can have a temperature dependency. In our simplified model, the quantity $\Delta^{(1)}(T)$  is directly calculated by:

\begin{equation}\label{gruneisen}
\Delta^{(1)}(T)=\omega_0\left[e^{-3\gamma \int_{0}^{T} \alpha(T') \,dT'}-1\right]
\end{equation}

where $\gamma$ is the Gr{\"u}neisen parameter, and $\alpha(T')$ is the coefficient of linear thermal expansion \cite{article}. In practice, we experimentally found values for $\omega(T)$ by measuring the tuning slope fitted it to  equation (\ref{dfdt}) with high accuracy.










When it comes to the wavelength dependent part of the index of refraction $n_{\lambda}\left(\lambda\right)$, we are basing our model in the most recent single-term Sellmeier equation found for synthetic diamond \cite{Turri:17}. The Sellmeier equation is usually calculated at room temperature (300~K), however the separable equation for the index of refraction in equation (\ref{eq:dispersion}) requires the index at absolute zero temperature. To that end, we approximated $n^0_{\lambda}\left(\lambda\right)$ as follows:

\begin{equation}
n^0_{\lambda}\left(\lambda\right)=n_{\lambda}(\lambda)- n_T\left(300\right)
\end{equation}

The factor $ n_T\left(300\right)$ was calculated using equation (\ref{eq:dispersion}) at 300~K. This way we also guarantee that the index of refraction is the one commonly known at room temperature.

In terms of the temperature dependence of the Raman shift, it defines the spectral range where the monolithic resonator will lase, although not the specific frequency of the Stokes standing waves. Having this in mind, we present here for completeness the dependency of the first order phonon frequency (or Raman shift) on temperature. The Klemens anharmonic approximation assumes that the zone-center phonon decays into two acoustical phonons of opposite momentum is appropriate to describe the effects in the diamond lattice \cite{PhysRev.148.845, PhysRevLett.75.1819}. In that model the relaxation time $\tau$ is

\begin{equation}
    \tau\simeq1+\frac{2}{e^{\frac{\hbar}{KT}\frac{\omega_R}{2}}-1}
\end{equation}


where $\omega_R$ is the Raman shift. The relaxation time $\tau$ is proportional to the Raman linewidth $\Delta \omega_R$ and that later is linearly connected to the Raman shift \cite{PhysRevB.61.3391}. The temperature-dependent Raman shift is then given by:

\begin{equation}
\omega_R(T)=1332.7-A_R\left(\frac{2}{e^{\frac{\hbar\omega_R}{2KT}}-1}\right)\frac{10^7}{c} \textrm{~[cm\textsuperscript{-1}]}
\end{equation}

where $A_R$ depends on the dispersion lines of diamond \cite{PhysRevB.61.3391}. A fit to the experimental data shown in  \cite{thesis} resulted in $A_R$~=~2.6~$\times$~10\textsuperscript{3}~GHz. We are now ready to experimentally measure $\partial\nu_S/\partial T$ and fit it to our model, allowing us to extract the thermo-optical coefficient of diamond.

\section{\label{sec:level1}Experiment\protect}

In order to confirm our method, we have setup an experiment to analyze the Stokes resonant frequency in the resonator as a function of temperature with high accuracy. From this data we can then relate the constants and tuning slopes related physical parameters. 


\begin{figure}[ht]
    \centering
    \includegraphics[width=8.6cm]{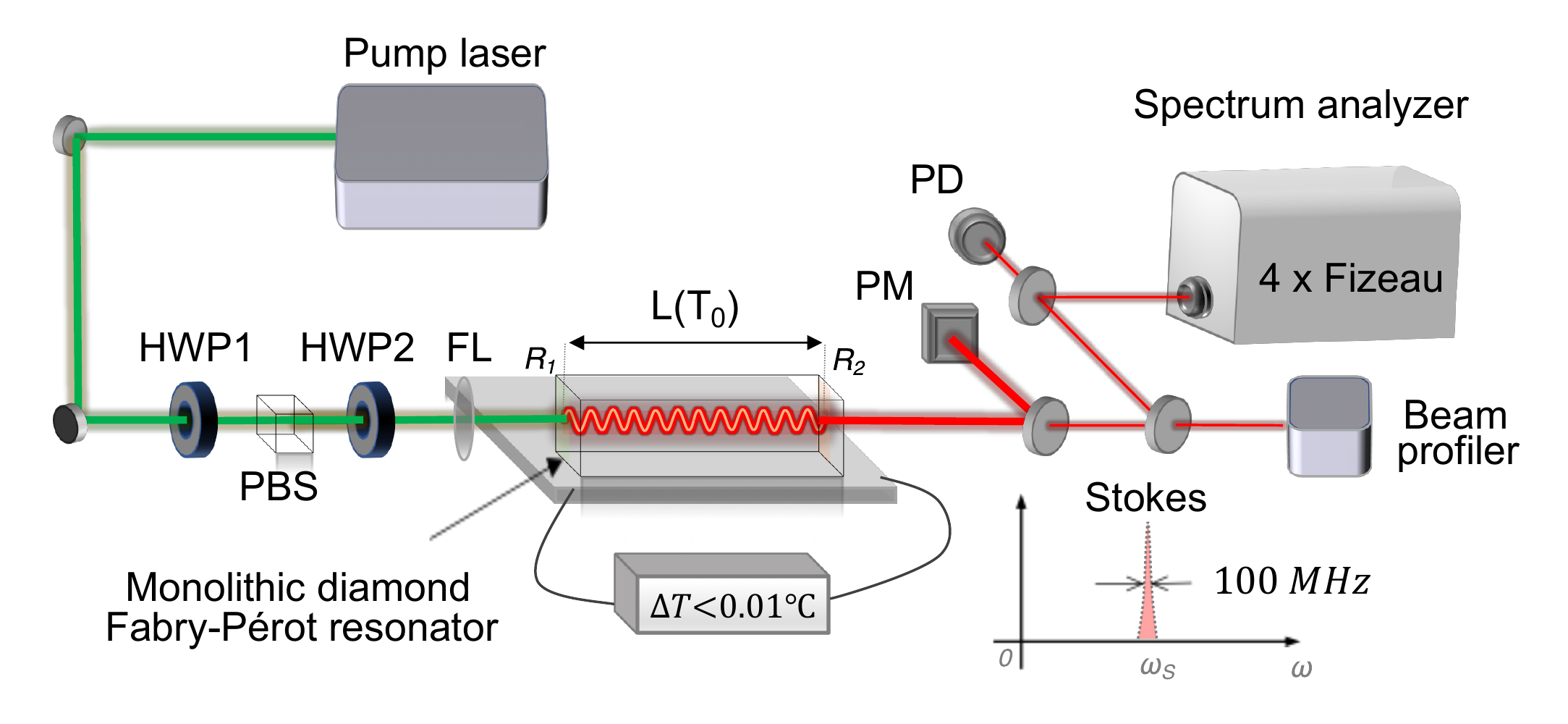}
    \caption{Schematic layout of the experimental setup: A monolithic diamond resonator is pumped by a frequency-doubled Q-switched Nd:YAG laser. The output Stokes nanosecond pulse was characterized temporally and spectrally with a set of four high resolution Fizeau interferometers, and a photodiode (PD) connected to a large bandwidth 16~GHz oscilloscope. HWP1, HWP2: half-wave-plates, PBS: polarizing beam splitter, FL: focusing lens, PM: power meter.}
    \label{fig:setup}
\end{figure}

A temperature adjustable monolithic FP diamond resonator was used in our experiments as the tool for measuring the optical properties of diamond under stimulated Raman scattering conditions. The Raman medium was a synthetic diamond cuboid crystal with dimensions of 7 $\times$ 2 $\times$ 2 mm\textsuperscript{3} (FSR @ 573 nm $\approx$ 8 GHz), plane-cut for beam propagation along the ⟨110⟩ axis and end-faces re-polished with a parallelism better than 0.5 \textmu m/mm. The experimental setup can be appreciated in Fig. \ref{fig:setup}.

Thanks to the high Raman gain of diamond at 532~nm, the Fresnel reflectivity of the un-coated surfaces ($R_1,R_2 \approx$ 18\%) was sufficient to ensure highly efficient Raman operation. The diamond crystal was placed on a copper mount inside a high precision oven (Covesion Ltd), with a temperature standard deviation of less than $<$10 mK. Note that the relatively small thermal expansion coefficient of diamond \cite{MOELLE1997839, JACOBSON2019107469} and dispersion \cite{index}, provided the necessary stability and robustness to perform our measurements accurately.

The pump is a frequency-doubled Nd:YAG 532~nm laser generating 10~ns pulses at a repetition rate of 100~Hz with an energy of 50~\textmu J. The pulses passed through a power control system consisting in a half-wave-plate (HWP1) and polarizing beam splitter (PBS). The polarization was controlled by means of another half-wave-plate (HPW2), note that the SRS process efficiency depends on polarization and is maximized when the pump polarization angle is parallel to the ⟨111⟩ crystallographic axis. The pump then arrives at the resonator and goes through the SRS process. 

\begin{figure}[htbp]
     \centering
     \includegraphics[width=8.6cm]{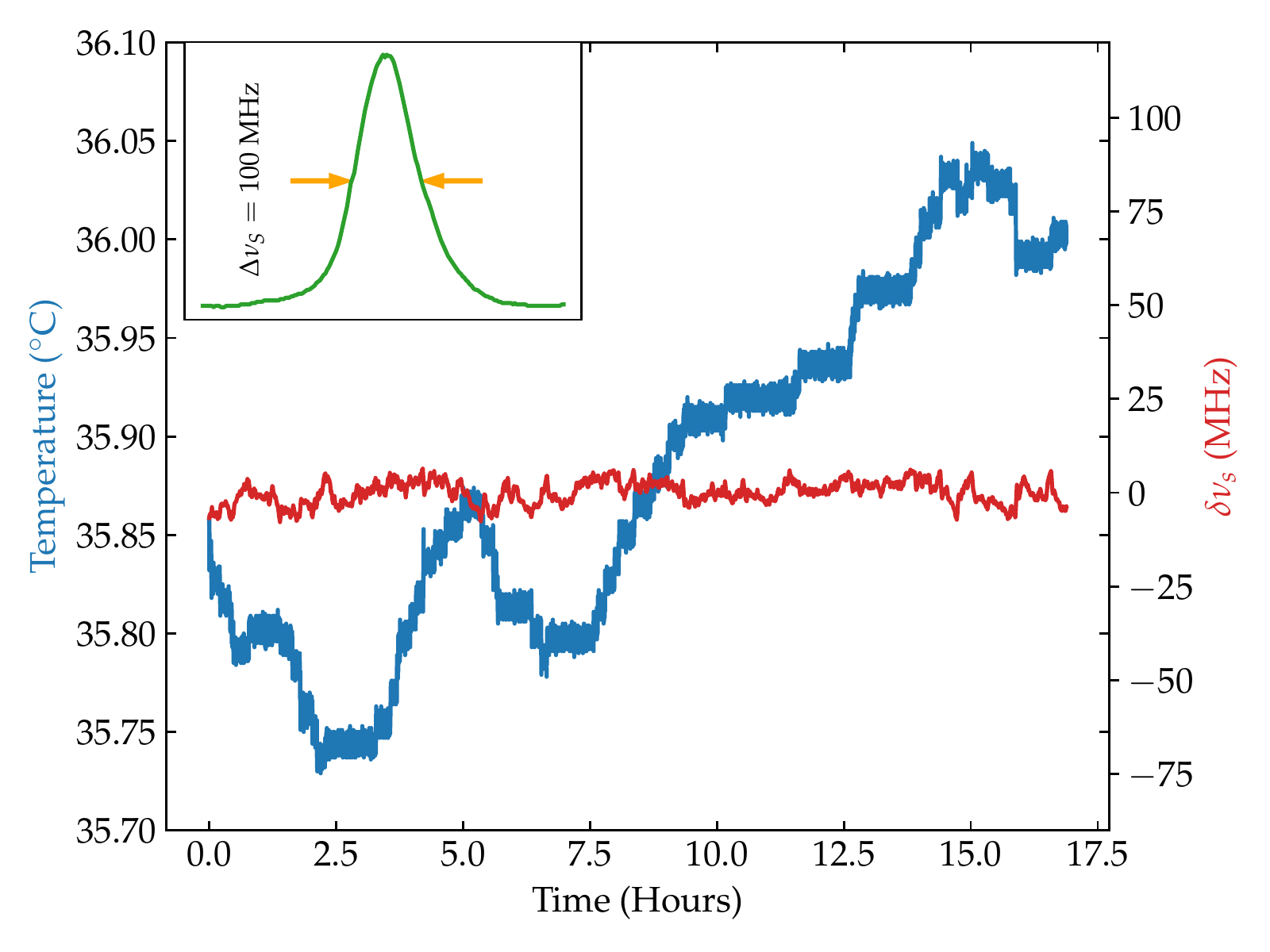}
     \caption{Active temperature stabilization of the Stokes resonant frequency over more than 16 hours. The RMS fluctuation of the Stokes frequency is less than 4~MHz. Inset: measured Stokes field spectrum.}
     \label{fig:stable}
 \end{figure}

The pump was focused into the diamond crystal by a 150~mm focal length lens (FL), producing a waist of 50~$\pm$~5~\textmu m in diameter and a resulting intensity of 0.1~GW/cm\textsuperscript{2}.  After the generation of the 1\textsuperscript{st} and 2\textsuperscript{nd} Stokes we used dichroics to filter the undesired Stokes orders. The resulting 573~nm beamn was then guided to the wavemeter, callibrated power meter (PM), photodiode (PD) and beam profiler. The linewidth (FWHM) of the 573~nm Stokes light was measured with a wavelength meter LM-007 wavemeter and was 100~$\pm$~20~MHz averaged over $\sim$1000 shots (shown inset in Fig. \ref{fig:stable}), whereas the center frequency deviation ($\delta \nu_S$) had an RMS value below $<$~4~MHz over more than 16~hours when actively stabilized using temperature as shown in Fig. \ref{fig:stable}.

\begin{figure}[htbp]
     \centering
     \includegraphics[width=8.6cm]{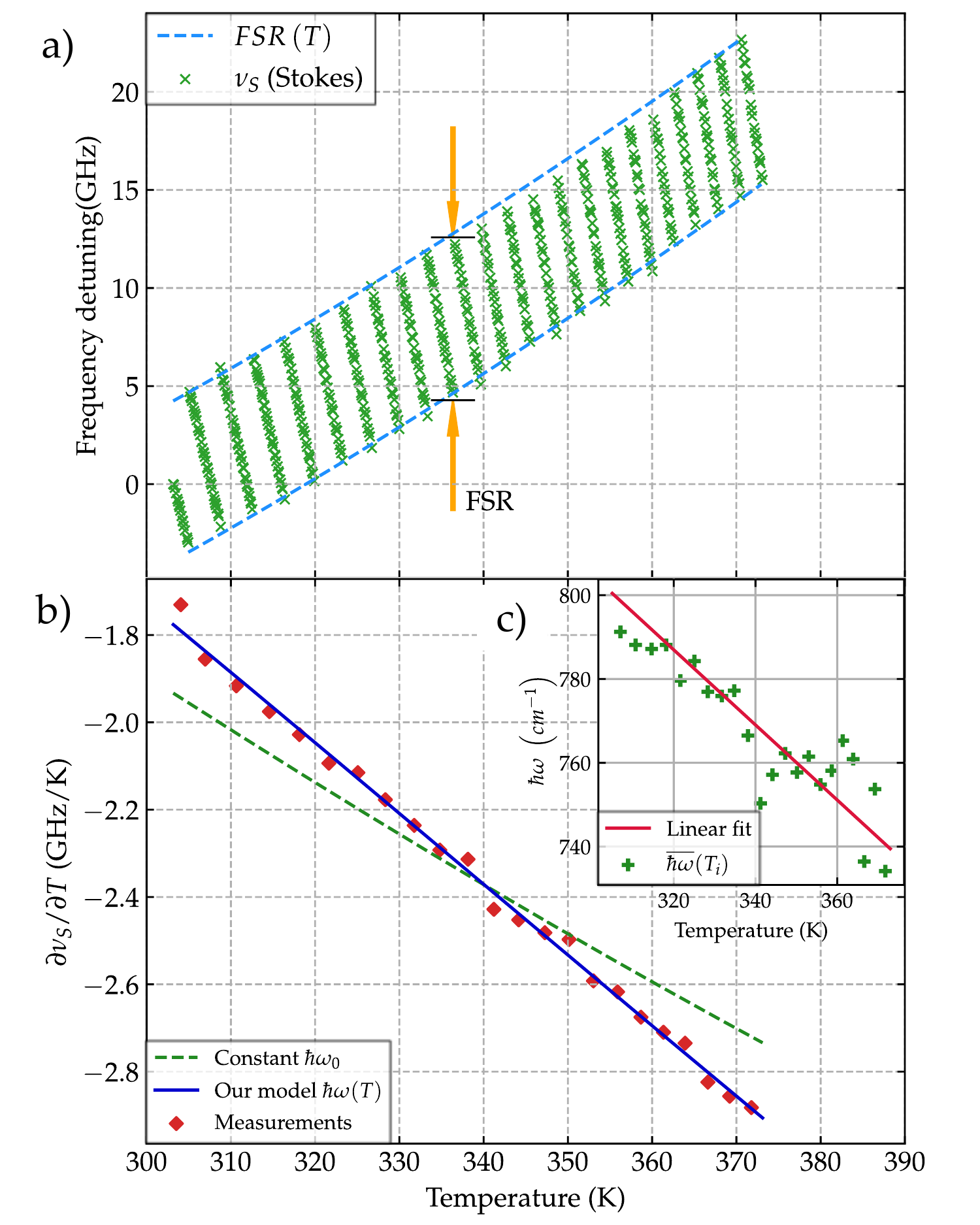}
     \caption{(a) Stokes center frequency ($\nu_S$) detuning as a function of measured diamond temperature. Dashed blue line represents the tuning range (FSR) of the Stokes frequency as a function of temperature. (b) Measured tuning slope for each FSR as a function of temperature. (dotted) Calculated tuning slope using the model for the thermo-optic coefficient in \cite{PhysRevB.62.16578} (dashed) Fit using the model presented in this paper using the temperature dependent  phonon frequencies. (c) Estimated average phonon frequencies as a function of temperature.}
     \label{fig:tuning}
 \end{figure}

The results of the measurement of the resonating Stokes wavelength with temperature are shown in Fig. \ref{fig:tuning} (a). The  tests were carried out by adjusting the temperature setting of the oven in increments of 10~mK. The average frequency-temperature tuning slope within a FSR of the resonator was approximately $\partial\nu_S/\partial T \approx$ -~2.3~GHz/K, whereas the temperature dependence of the first-order Raman phonon line was about $\partial\nu_R/\partial T \approx$ +0.23 GHz/K. This agrees reasonably with calculations resulting from the Klemens model ($\approx$ +0.2 to 0.25 GHz/K  between 300 and 400~K). 

Figure \ref{fig:tuning} (b) shows the measured slope in each FSR as a function of temperature. It can be appreciated that the overall tuning slope increases in absolute value as a function of temperature due to the temperature dependency of the thermo-optic coefficient. The slope in the tuning curves varies significantly from -1.8~GHz/K to -2.8~GHz/K in about 70~K. The experimental results are compared with the model in \cite{PhysRevB.62.16578} (dashed green line, constant average phonon frequency $\hbar\omega_0 \approx$ 711~cm\textsuperscript{-1}) and the model presented in this work (solid blue, resulting from using the estimated $\omega(T)$ instead).

 We used the values of the slope to estimate then the average phonon frequency $\hbar\omega(T)$ of each FSR by fitting the data to equation (\ref{dfdt}) infering the phonon frequency as in equation (\ref{phonon_energy}). Figure \ref{fig:tuning} (c) shows the fitted values for the phonon frequency $\hbar\omega(T)$ for each FSR. From these results we can see that the average phonon frequency under strong SRS has a linear decreasing dependency with temperature. The linear extrapolation shows a rate of change of $-0.9~\pm~0.05$~cm\textsuperscript{-1}/K  at the given temperature range, starting at 805~$\pm$~32~cm\textsuperscript{-1} at 300~K down to 740 ~cm\textsuperscript{-1} at 370~K.

Compare the fitted linearly decreasing tendency of the phonon frequency with temperature with the predictions calculated using  equation (\ref{gruneisen}). As it can be inferred, the dependency should be relatively linear for small temperature increments, where the lattice expands uniformly in all directions and the thermal expansion coefficient is nearly constant. The resulting phonon frequency then linearly decays with temperature, which is expected by approximating $\omega_0 e^{-3\gamma\alpha T} \approx \omega_0(1-3\gamma\alpha T)$. From this trend it is then possible to estimate the Gr{\"u}neisen parameter for diamond $\gamma$~=~4715 and $\hbar\omega_0=600$~cm\textsuperscript{-1}.

 \begin{figure}[h]
    \centering
    \includegraphics[width=8.6cm]{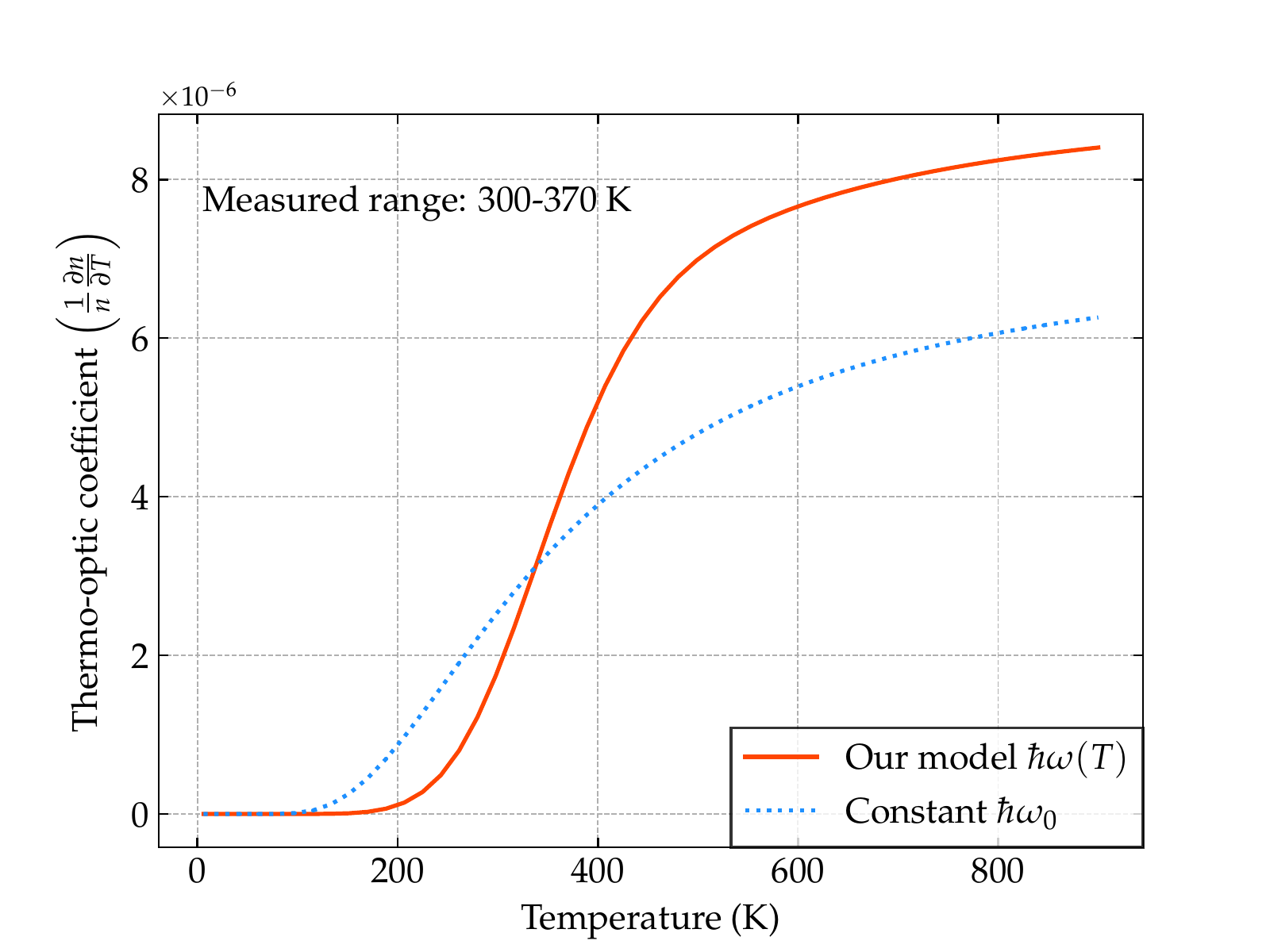}
    \caption{Calculated thermo-optic coefficient of diamond as a function of temperature between 0 -- 900~K using the model presented in this work (red solid), and the model presented in \cite{PhysRevB.62.16578} (dotted blue). The experimental measured range corresponds to 300 -- 370~K.}
    \label{fig:fit}
\end{figure}

We can now proceed to estimate the thermo-optic coefficient directly by derivating equation (\ref{nt}) with the values measured for the temperature-dependent average phonon frequency $\hbar\omega(T)$. The result of this extrapolation is shown in Fig. \ref{fig:fit} (solid red), alongside with the previously estimated values found in \cite{PhysRevB.62.16578} (dashed blue). 

It is clear that for the overall thermo-optic coefficient at room temperature is approximately 3.5~$\times$~10\textsuperscript{-6} for both cases, however, for our model the second moment of the index of refraction $\partial^2 n/\partial T^2$ with respect to the temperature is appreciably larger. Interestingly, the range where the index of refraction is nonlinear is most severe for temperatures in the range from 200--400~K. Below 200~K, $(1/n) \partial n / \partial T$ is nearly zero, whereas for values above 400~K it asymptotically tends to 8$\times$10\textsuperscript{-6}~K\textsuperscript{-1}.\\

\section{Conclusions}

In this work we studied the rate of change of the resonant Stokes wavelength inside a monolithic Raman resonator with temperature. We found that existing models for the temperature dependency of the refractive index correspond approximately to observed experimental processes, however the accuracy in their predictions is poorer in the 300~--~400~K range. Here, since $\partial \nu_S/\partial T$ depends directly on the thermo-optic coefficient, we proposed to scan the temperature while measuring the resonant Stokes wavelength to re-calculate the thermal dependency of the index of refraction. We call this approach "the resonant Stokes field" method.

Regarding the flexibility of the proposed method, the combination of very narrow spectral bandwidth and resulting high spectral density from the resonator, alongside with the large transparency range of diamond, make it very versatile and useful at a large range of wavelengths and temperatures. In fact, the modest requirements in terms of resonator quality factors readily allow for stable and portable operation readily usable in scientific applications.

Furthermore, we propose a model for estimating the average lattice phonon frequency based on the thermal expansion and the Gr{\"u}neisen parameter. We showed that for small temperature increments the  dependency of average phonon frequency with temperature is linearly decreasing. We expect that the presented method and measured diamond thermo-optical parameters will be useful for research related to the development of temperature-sensitive integrated photonic devices in diamond.


\bibliography{apssamp}

\providecommand{\noopsort}[1]{}\providecommand{\singleletter}[1]{#1}%
\begin{thebibliography}{41}%
\makeatletter
\providecommand \@ifxundefined [1]{%
 \@ifx{#1\undefined}
}%
\providecommand \@ifnum [1]{%
 \ifnum #1\expandafter \@firstoftwo
 \else \expandafter \@secondoftwo
 \fi
}%
\providecommand \@ifx [1]{%
 \ifx #1\expandafter \@firstoftwo
 \else \expandafter \@secondoftwo
 \fi
}%
\providecommand \natexlab [1]{#1}%
\providecommand \enquote  [1]{``#1''}%
\providecommand \bibnamefont  [1]{#1}%
\providecommand \bibfnamefont [1]{#1}%
\providecommand \citenamefont [1]{#1}%
\providecommand \href@noop [0]{\@secondoftwo}%
\providecommand \href [0]{\begingroup \@sanitize@url \@href}%
\providecommand \@href[1]{\@@startlink{#1}\@@href}%
\providecommand \@@href[1]{\endgroup#1\@@endlink}%
\providecommand \@sanitize@url [0]{\catcode `\\12\catcode `\$12\catcode
  `\&12\catcode `\#12\catcode `\^12\catcode `\_12\catcode `\%12\relax}%
\providecommand \@@startlink[1]{}%
\providecommand \@@endlink[0]{}%
\providecommand \url  [0]{\begingroup\@sanitize@url \@url }%
\providecommand \@url [1]{\endgroup\@href {#1}{\urlprefix }}%
\providecommand \urlprefix  [0]{URL }%
\providecommand \Eprint [0]{\href }%
\providecommand \doibase [0]{https://doi.org/}%
\providecommand \selectlanguage [0]{\@gobble}%
\providecommand \bibinfo  [0]{\@secondoftwo}%
\providecommand \bibfield  [0]{\@secondoftwo}%
\providecommand \translation [1]{[#1]}%
\providecommand \BibitemOpen [0]{}%
\providecommand \bibitemStop [0]{}%
\providecommand \bibitemNoStop [0]{.\EOS\space}%
\providecommand \EOS [0]{\spacefactor3000\relax}%
\providecommand \BibitemShut  [1]{\csname bibitem#1\endcsname}%
\let\auto@bib@innerbib\@empty
\bibitem [{\citenamefont
  {Mildren}()}]{doi:https://doi.org/10.1002/9783527648603.ch1}%
  \BibitemOpen
  \bibfield  {author} {\bibinfo {author} {\bibfnamefont {R.~P.}\ \bibnamefont
  {Mildren}},\ }\bibinfo {title} {Intrinsic optical properties of diamond},\
  in\ \href@noop {} {\emph {\bibinfo {booktitle} {Optical Engineering of
  Diamond}}}\ (\bibinfo  {publisher} {John Wiley \& Sons, Ltd})\ Chap.~\bibinfo
  {chapter} {1}, pp.\ \bibinfo {pages} {1--34}\BibitemShut {NoStop}%
\bibitem [{\citenamefont {Teraji}(2020)}]{TERAJI202037}%
  \BibitemOpen
  \bibfield  {author} {\bibinfo {author} {\bibfnamefont {T.}~\bibnamefont
  {Teraji}},\ }\bibfield  {title} {\bibinfo {title} {Chapter two - ultrapure
  homoepitaxial diamond films grown by chemical vapor deposition for quantum
  device application},\ }in\ \href
  {https://doi.org/https://doi.org/10.1016/bs.semsem.2020.03.002} {\emph
  {\bibinfo {booktitle} {Diamond for Quantum Applications Part 1}}},\ \bibinfo
  {series} {Semiconductors and Semimetals}, Vol.\ \bibinfo {volume} {103},\
  \bibinfo {editor} {edited by\ \bibinfo {editor} {\bibfnamefont {C.~E.}\
  \bibnamefont {Nebel}}, \bibinfo {editor} {\bibfnamefont {I.}~\bibnamefont
  {Aharonovich}}, \bibinfo {editor} {\bibfnamefont {N.}~\bibnamefont
  {Mizuochi}},\ and\ \bibinfo {editor} {\bibfnamefont {M.}~\bibnamefont
  {Hatano}}}\ (\bibinfo  {publisher} {Elsevier},\ \bibinfo {year} {2020})\ pp.\
  \bibinfo {pages} {37--55}\BibitemShut {NoStop}%
\bibitem [{\citenamefont {Neu}\ and\ \citenamefont
  {Becher}(2014)}]{NEU2014127}%
  \BibitemOpen
  \bibfield  {author} {\bibinfo {author} {\bibfnamefont {E.}~\bibnamefont
  {Neu}}\ and\ \bibinfo {author} {\bibfnamefont {C.}~\bibnamefont {Becher}},\
  }\bibfield  {title} {\bibinfo {title} {6 - diamond-based single-photon
  sources and their application in quantum key distribution},\ }in\ \href
  {https://doi.org/https://doi.org/10.1533/9780857096685.2.127} {\emph
  {\bibinfo {booktitle} {Quantum Information Processing with Diamond}}},\
  \bibinfo {editor} {edited by\ \bibinfo {editor} {\bibfnamefont
  {S.}~\bibnamefont {Prawer}}\ and\ \bibinfo {editor} {\bibfnamefont
  {I.}~\bibnamefont {Aharonovich}}}\ (\bibinfo  {publisher} {Woodhead
  Publishing},\ \bibinfo {year} {2014})\ pp.\ \bibinfo {pages}
  {127--159}\BibitemShut {NoStop}%
\bibitem [{\citenamefont {Markham}\ \emph {et~al.}(2019)\citenamefont
  {Markham}, \citenamefont {Edmonds}, \citenamefont {Bennett}, \citenamefont
  {Colard}, \citenamefont {Hillman},\ and\ \citenamefont
  {Jaszczykowski}}]{Markham:19}%
  \BibitemOpen
  \bibfield  {author} {\bibinfo {author} {\bibfnamefont {M.}~\bibnamefont
  {Markham}}, \bibinfo {author} {\bibfnamefont {A.}~\bibnamefont {Edmonds}},
  \bibinfo {author} {\bibfnamefont {A.}~\bibnamefont {Bennett}}, \bibinfo
  {author} {\bibfnamefont {P.-O.}\ \bibnamefont {Colard}}, \bibinfo {author}
  {\bibfnamefont {W.}~\bibnamefont {Hillman}},\ and\ \bibinfo {author}
  {\bibfnamefont {M.}~\bibnamefont {Jaszczykowski}},\ }\bibfield  {title}
  {\bibinfo {title} {Cvd diamond for quantum applications},\ }in\ \href
  {https://doi.org/10.1364/DP.2019.135} {\emph {\bibinfo {booktitle} {Symposium
  Latsis 2019 on Diamond Photonics - Physics, Technologies and Applications}}}\
  (\bibinfo  {publisher} {Optical Society of America},\ \bibinfo {year}
  {2019})\ p.\ \bibinfo {pages} {135}\BibitemShut {NoStop}%
\bibitem [{\citenamefont {Sukachev}\ \emph {et~al.}(2017)\citenamefont
  {Sukachev}, \citenamefont {Sipahigil}, \citenamefont {Nguyen}, \citenamefont
  {Bhaskar}, \citenamefont {Evans}, \citenamefont {Jelezko},\ and\
  \citenamefont {Lukin}}]{PhysRevLett.119.223602}%
  \BibitemOpen
  \bibfield  {author} {\bibinfo {author} {\bibfnamefont {D.~D.}\ \bibnamefont
  {Sukachev}}, \bibinfo {author} {\bibfnamefont {A.}~\bibnamefont {Sipahigil}},
  \bibinfo {author} {\bibfnamefont {C.~T.}\ \bibnamefont {Nguyen}}, \bibinfo
  {author} {\bibfnamefont {M.~K.}\ \bibnamefont {Bhaskar}}, \bibinfo {author}
  {\bibfnamefont {R.~E.}\ \bibnamefont {Evans}}, \bibinfo {author}
  {\bibfnamefont {F.}~\bibnamefont {Jelezko}},\ and\ \bibinfo {author}
  {\bibfnamefont {M.~D.}\ \bibnamefont {Lukin}},\ }\bibfield  {title} {\bibinfo
  {title} {Silicon-vacancy spin qubit in diamond: A quantum memory exceeding 10
  ms with single-shot state readout},\ }\href
  {https://doi.org/10.1103/PhysRevLett.119.223602} {\bibfield  {journal}
  {\bibinfo  {journal} {Phys. Rev. Lett.}\ }\textbf {\bibinfo {volume} {119}},\
  \bibinfo {pages} {223602} (\bibinfo {year} {2017})}\BibitemShut {NoStop}%
\bibitem [{\citenamefont {Fontanella}\ \emph {et~al.}(1977)\citenamefont
  {Fontanella}, \citenamefont {Johnston}, \citenamefont {Colwell},\ and\
  \citenamefont {Andeen}}]{Fontanella:77}%
  \BibitemOpen
  \bibfield  {author} {\bibinfo {author} {\bibfnamefont {J.}~\bibnamefont
  {Fontanella}}, \bibinfo {author} {\bibfnamefont {R.~L.}\ \bibnamefont
  {Johnston}}, \bibinfo {author} {\bibfnamefont {J.~H.}\ \bibnamefont
  {Colwell}},\ and\ \bibinfo {author} {\bibfnamefont {C.}~\bibnamefont
  {Andeen}},\ }\bibfield  {title} {\bibinfo {title} {Temperature and pressure
  variation of the refractive index of diamond},\ }\href
  {https://doi.org/10.1364/AO.16.002949} {\bibfield  {journal} {\bibinfo
  {journal} {Appl. Opt.}\ }\textbf {\bibinfo {volume} {16}},\ \bibinfo {pages}
  {2949} (\bibinfo {year} {1977})}\BibitemShut {NoStop}%
\bibitem [{\citenamefont {Ruf}\ \emph {et~al.}(2000)\citenamefont {Ruf},
  \citenamefont {Cardona}, \citenamefont {Pickles},\ and\ \citenamefont
  {Sussmann}}]{PhysRevB.62.16578}%
  \BibitemOpen
  \bibfield  {author} {\bibinfo {author} {\bibfnamefont {T.}~\bibnamefont
  {Ruf}}, \bibinfo {author} {\bibfnamefont {M.}~\bibnamefont {Cardona}},
  \bibinfo {author} {\bibfnamefont {C.~S.~J.}\ \bibnamefont {Pickles}},\ and\
  \bibinfo {author} {\bibfnamefont {R.}~\bibnamefont {Sussmann}},\ }\bibfield
  {title} {\bibinfo {title} {Temperature dependence of the refractive index of
  diamond up to $925 \mathrm{K}$},\ }\href
  {https://doi.org/10.1103/PhysRevB.62.16578} {\bibfield  {journal} {\bibinfo
  {journal} {Phys. Rev. B}\ }\textbf {\bibinfo {volume} {62}},\ \bibinfo
  {pages} {16578} (\bibinfo {year} {2000})}\BibitemShut {NoStop}%
\bibitem [{\citenamefont {Yu}\ and\ \citenamefont
  {Cardona}(1970)}]{PhysRevB.2.3193}%
  \BibitemOpen
  \bibfield  {author} {\bibinfo {author} {\bibfnamefont {P.~Y.}\ \bibnamefont
  {Yu}}\ and\ \bibinfo {author} {\bibfnamefont {M.}~\bibnamefont {Cardona}},\
  }\bibfield  {title} {\bibinfo {title} {Temperature coefficient of the
  refractive index of diamond- and zinc-blende-type semiconductors},\ }\href
  {https://doi.org/10.1103/PhysRevB.2.3193} {\bibfield  {journal} {\bibinfo
  {journal} {Phys. Rev. B}\ }\textbf {\bibinfo {volume} {2}},\ \bibinfo {pages}
  {3193} (\bibinfo {year} {1970})}\BibitemShut {NoStop}%
\bibitem [{\citenamefont {Lux}\ \emph {et~al.}(2016{\natexlab{a}})\citenamefont
  {Lux}, \citenamefont {Sarang}, \citenamefont {Kitzler}, \citenamefont
  {Spence},\ and\ \citenamefont {Mildren}}]{Lux:16}%
  \BibitemOpen
  \bibfield  {author} {\bibinfo {author} {\bibfnamefont {O.}~\bibnamefont
  {Lux}}, \bibinfo {author} {\bibfnamefont {S.}~\bibnamefont {Sarang}},
  \bibinfo {author} {\bibfnamefont {O.}~\bibnamefont {Kitzler}}, \bibinfo
  {author} {\bibfnamefont {D.~J.}\ \bibnamefont {Spence}},\ and\ \bibinfo
  {author} {\bibfnamefont {R.~P.}\ \bibnamefont {Mildren}},\ }\bibfield
  {title} {\bibinfo {title} {Intrinsically stable high-power single
  longitudinal mode laser using spatial hole burning free gain},\ }\href
  {https://doi.org/10.1364/OPTICA.3.000876} {\bibfield  {journal} {\bibinfo
  {journal} {Optica}\ }\textbf {\bibinfo {volume} {3}},\ \bibinfo {pages} {876}
  (\bibinfo {year} {2016}{\natexlab{a}})}\BibitemShut {NoStop}%
\bibitem [{\citenamefont {Granados}\ \emph {et~al.}(2011)\citenamefont
  {Granados}, \citenamefont {Spence},\ and\ \citenamefont
  {Mildren}}]{Granados:11}%
  \BibitemOpen
  \bibfield  {author} {\bibinfo {author} {\bibfnamefont {E.}~\bibnamefont
  {Granados}}, \bibinfo {author} {\bibfnamefont {D.~J.}\ \bibnamefont
  {Spence}},\ and\ \bibinfo {author} {\bibfnamefont {R.~P.}\ \bibnamefont
  {Mildren}},\ }\bibfield  {title} {\bibinfo {title} {Deep ultraviolet diamond
  raman laser},\ }\href {https://doi.org/10.1364/OE.19.010857} {\bibfield
  {journal} {\bibinfo  {journal} {Opt. Express}\ }\textbf {\bibinfo {volume}
  {19}},\ \bibinfo {pages} {10857} (\bibinfo {year} {2011})}\BibitemShut
  {NoStop}%
\bibitem [{\citenamefont {Mildren}\ and\ \citenamefont
  {Sabella}(2009)}]{Mildren:09}%
  \BibitemOpen
  \bibfield  {author} {\bibinfo {author} {\bibfnamefont {R.~P.}\ \bibnamefont
  {Mildren}}\ and\ \bibinfo {author} {\bibfnamefont {A.}~\bibnamefont
  {Sabella}},\ }\bibfield  {title} {\bibinfo {title} {Highly efficient diamond
  raman laser},\ }\href {https://doi.org/10.1364/OL.34.002811} {\bibfield
  {journal} {\bibinfo  {journal} {Opt. Lett.}\ }\textbf {\bibinfo {volume}
  {34}},\ \bibinfo {pages} {2811} (\bibinfo {year} {2009})}\BibitemShut
  {NoStop}%
\bibitem [{\citenamefont {Spence}\ \emph {et~al.}(2010)\citenamefont {Spence},
  \citenamefont {Granados},\ and\ \citenamefont {Mildren}}]{Spence:10}%
  \BibitemOpen
  \bibfield  {author} {\bibinfo {author} {\bibfnamefont {D.~J.}\ \bibnamefont
  {Spence}}, \bibinfo {author} {\bibfnamefont {E.}~\bibnamefont {Granados}},\
  and\ \bibinfo {author} {\bibfnamefont {R.~P.}\ \bibnamefont {Mildren}},\
  }\bibfield  {title} {\bibinfo {title} {Mode-locked picosecond diamond raman
  laser},\ }\href {https://doi.org/10.1364/OL.35.000556} {\bibfield  {journal}
  {\bibinfo  {journal} {Opt. Lett.}\ }\textbf {\bibinfo {volume} {35}},\
  \bibinfo {pages} {556} (\bibinfo {year} {2010})}\BibitemShut {NoStop}%
\bibitem [{\citenamefont {Lubeigt}\ \emph {et~al.}(2010)\citenamefont
  {Lubeigt}, \citenamefont {Bonner}, \citenamefont {Hastie}, \citenamefont
  {Dawson}, \citenamefont {Burns},\ and\ \citenamefont {Kemp}}]{Lubeigt:10}%
  \BibitemOpen
  \bibfield  {author} {\bibinfo {author} {\bibfnamefont {W.}~\bibnamefont
  {Lubeigt}}, \bibinfo {author} {\bibfnamefont {G.~M.}\ \bibnamefont {Bonner}},
  \bibinfo {author} {\bibfnamefont {J.~E.}\ \bibnamefont {Hastie}}, \bibinfo
  {author} {\bibfnamefont {M.~D.}\ \bibnamefont {Dawson}}, \bibinfo {author}
  {\bibfnamefont {D.}~\bibnamefont {Burns}},\ and\ \bibinfo {author}
  {\bibfnamefont {A.~J.}\ \bibnamefont {Kemp}},\ }\bibfield  {title} {\bibinfo
  {title} {Continuous-wave diamond raman laser},\ }\href
  {https://doi.org/10.1364/OL.35.002994} {\bibfield  {journal} {\bibinfo
  {journal} {Opt. Lett.}\ }\textbf {\bibinfo {volume} {35}},\ \bibinfo {pages}
  {2994} (\bibinfo {year} {2010})}\BibitemShut {NoStop}%
\bibitem [{\citenamefont {Sabella}\ \emph {et~al.}(2014)\citenamefont
  {Sabella}, \citenamefont {Piper},\ and\ \citenamefont
  {Mildren}}]{Sabella:14}%
  \BibitemOpen
  \bibfield  {author} {\bibinfo {author} {\bibfnamefont {A.}~\bibnamefont
  {Sabella}}, \bibinfo {author} {\bibfnamefont {J.~A.}\ \bibnamefont {Piper}},\
  and\ \bibinfo {author} {\bibfnamefont {R.~P.}\ \bibnamefont {Mildren}},\
  }\bibfield  {title} {\bibinfo {title} {Diamond raman laser with continuously
  tunable output from 3.38 to 3.80~\textmu m},\ }\href
  {https://doi.org/10.1364/OL.39.004037} {\bibfield  {journal} {\bibinfo
  {journal} {Opt. Lett.}\ }\textbf {\bibinfo {volume} {39}},\ \bibinfo {pages}
  {4037} (\bibinfo {year} {2014})}\BibitemShut {NoStop}%
\bibitem [{\citenamefont {Echarri}\ \emph {et~al.}(2020)\citenamefont
  {Echarri}, \citenamefont {Chrysalidis}, \citenamefont {Fedosseev},
  \citenamefont {Marsh}, \citenamefont {Mildren}, \citenamefont {Olaizola},
  \citenamefont {Spence}, \citenamefont {Wilkins},\ and\ \citenamefont
  {Granados}}]{Echarri:20}%
  \BibitemOpen
  \bibfield  {author} {\bibinfo {author} {\bibfnamefont {D.~T.}\ \bibnamefont
  {Echarri}}, \bibinfo {author} {\bibfnamefont {K.}~\bibnamefont
  {Chrysalidis}}, \bibinfo {author} {\bibfnamefont {V.~N.}\ \bibnamefont
  {Fedosseev}}, \bibinfo {author} {\bibfnamefont {B.~A.}\ \bibnamefont
  {Marsh}}, \bibinfo {author} {\bibfnamefont {R.~P.}\ \bibnamefont {Mildren}},
  \bibinfo {author} {\bibfnamefont {S.~M.}\ \bibnamefont {Olaizola}}, \bibinfo
  {author} {\bibfnamefont {D.~J.}\ \bibnamefont {Spence}}, \bibinfo {author}
  {\bibfnamefont {S.~G.}\ \bibnamefont {Wilkins}},\ and\ \bibinfo {author}
  {\bibfnamefont {E.}~\bibnamefont {Granados}},\ }\bibfield  {title} {\bibinfo
  {title} {Broadly tunable linewidth-invariant raman stokes comb for selective
  resonance photoionization},\ }\href {https://doi.org/10.1364/OE.384630}
  {\bibfield  {journal} {\bibinfo  {journal} {Opt. Express}\ }\textbf {\bibinfo
  {volume} {28}},\ \bibinfo {pages} {8589} (\bibinfo {year}
  {2020})}\BibitemShut {NoStop}%
\bibitem [{\citenamefont {Friel}\ \emph {et~al.}(2010)\citenamefont {Friel},
  \citenamefont {Geoghegan}, \citenamefont {Twitchen},\ and\ \citenamefont
  {Scarsbrook}}]{10.1117/12.864981}%
  \BibitemOpen
  \bibfield  {author} {\bibinfo {author} {\bibfnamefont {I.}~\bibnamefont
  {Friel}}, \bibinfo {author} {\bibfnamefont {S.~L.}\ \bibnamefont
  {Geoghegan}}, \bibinfo {author} {\bibfnamefont {D.~J.}\ \bibnamefont
  {Twitchen}},\ and\ \bibinfo {author} {\bibfnamefont {G.~A.}\ \bibnamefont
  {Scarsbrook}},\ }\bibfield  {title} {\bibinfo {title} {{Development of high
  quality single crystal diamond for novel laser applications}},\ }in\ \href
  {https://doi.org/10.1117/12.864981} {\emph {\bibinfo {booktitle} {Optics and
  Photonics for Counterterrorism and Crime Fighting VI and Optical Materials in
  Defence Systems Technology VII}}},\ Vol.\ \bibinfo {volume} {7838},\ \bibinfo
  {editor} {edited by\ \bibinfo {editor} {\bibfnamefont {C.}~\bibnamefont
  {Lewis}}, \bibinfo {editor} {\bibfnamefont {D.}~\bibnamefont {Burgess}},
  \bibinfo {editor} {\bibfnamefont {R.}~\bibnamefont {Zamboni}}, \bibinfo
  {editor} {\bibfnamefont {F.}~\bibnamefont {Kajzar}},\ and\ \bibinfo {editor}
  {\bibfnamefont {E.~M.}\ \bibnamefont {Heckman}}},\ \bibinfo {organization}
  {International Society for Optics and Photonics}\ (\bibinfo  {publisher}
  {SPIE},\ \bibinfo {year} {2010})\ pp.\ \bibinfo {pages} {340 --
  347}\BibitemShut {NoStop}%
\bibitem [{\citenamefont {Antipov}\ \emph {et~al.}(2019)\citenamefont
  {Antipov}, \citenamefont {Sabella}, \citenamefont {Williams}, \citenamefont
  {Kitzler}, \citenamefont {Spence},\ and\ \citenamefont
  {Mildren}}]{Antipov:19}%
  \BibitemOpen
  \bibfield  {author} {\bibinfo {author} {\bibfnamefont {S.}~\bibnamefont
  {Antipov}}, \bibinfo {author} {\bibfnamefont {A.}~\bibnamefont {Sabella}},
  \bibinfo {author} {\bibfnamefont {R.~J.}\ \bibnamefont {Williams}}, \bibinfo
  {author} {\bibfnamefont {O.}~\bibnamefont {Kitzler}}, \bibinfo {author}
  {\bibfnamefont {D.~J.}\ \bibnamefont {Spence}},\ and\ \bibinfo {author}
  {\bibfnamefont {R.~P.}\ \bibnamefont {Mildren}},\ }\bibfield  {title}
  {\bibinfo {title} {1.2 kw quasi-steady-state diamond raman laser pumped by an
  $m^2=15$ beam},\ }\href {https://doi.org/10.1364/OL.44.002506} {\bibfield
  {journal} {\bibinfo  {journal} {Opt. Lett.}\ }\textbf {\bibinfo {volume}
  {44}},\ \bibinfo {pages} {2506} (\bibinfo {year} {2019})}\BibitemShut
  {NoStop}%
\bibitem [{\citenamefont {Lux}\ \emph {et~al.}(2016{\natexlab{b}})\citenamefont
  {Lux}, \citenamefont {Sarang}, \citenamefont {Williams}, \citenamefont
  {McKay},\ and\ \citenamefont {Mildren}}]{Lux:16b}%
  \BibitemOpen
  \bibfield  {author} {\bibinfo {author} {\bibfnamefont {O.}~\bibnamefont
  {Lux}}, \bibinfo {author} {\bibfnamefont {S.}~\bibnamefont {Sarang}},
  \bibinfo {author} {\bibfnamefont {R.~J.}\ \bibnamefont {Williams}}, \bibinfo
  {author} {\bibfnamefont {A.}~\bibnamefont {McKay}},\ and\ \bibinfo {author}
  {\bibfnamefont {R.~P.}\ \bibnamefont {Mildren}},\ }\bibfield  {title}
  {\bibinfo {title} {Single longitudinal mode diamond raman laser in the
  eye-safe spectral region for water vapor detection},\ }\href
  {https://doi.org/10.1364/OE.24.027812} {\bibfield  {journal} {\bibinfo
  {journal} {Opt. Express}\ }\textbf {\bibinfo {volume} {24}},\ \bibinfo
  {pages} {27812} (\bibinfo {year} {2016}{\natexlab{b}})}\BibitemShut {NoStop}%
\bibitem [{\citenamefont {Kitzler}\ \emph {et~al.}(2017)\citenamefont
  {Kitzler}, \citenamefont {Lin}, \citenamefont {Pask}, \citenamefont
  {Mildren}, \citenamefont {Webster}, \citenamefont {Hempler}, \citenamefont
  {Malcolm},\ and\ \citenamefont {Spence}}]{Kitzler:17}%
  \BibitemOpen
  \bibfield  {author} {\bibinfo {author} {\bibfnamefont {O.}~\bibnamefont
  {Kitzler}}, \bibinfo {author} {\bibfnamefont {J.}~\bibnamefont {Lin}},
  \bibinfo {author} {\bibfnamefont {H.~M.}\ \bibnamefont {Pask}}, \bibinfo
  {author} {\bibfnamefont {R.~P.}\ \bibnamefont {Mildren}}, \bibinfo {author}
  {\bibfnamefont {S.~C.}\ \bibnamefont {Webster}}, \bibinfo {author}
  {\bibfnamefont {N.}~\bibnamefont {Hempler}}, \bibinfo {author} {\bibfnamefont
  {G.~P.~A.}\ \bibnamefont {Malcolm}},\ and\ \bibinfo {author} {\bibfnamefont
  {D.~J.}\ \bibnamefont {Spence}},\ }\bibfield  {title} {\bibinfo {title}
  {Single-longitudinal-mode ring diamond raman laser},\ }\href
  {https://doi.org/10.1364/OL.42.001229} {\bibfield  {journal} {\bibinfo
  {journal} {Opt. Lett.}\ }\textbf {\bibinfo {volume} {42}},\ \bibinfo {pages}
  {1229} (\bibinfo {year} {2017})}\BibitemShut {NoStop}%
\bibitem [{\citenamefont {Sarang}\ \emph {et~al.}(2019)\citenamefont {Sarang},
  \citenamefont {Kitzler}, \citenamefont {Lux}, \citenamefont {Bai},
  \citenamefont {Williams}, \citenamefont {Spence},\ and\ \citenamefont
  {Mildren}}]{Sarang:19}%
  \BibitemOpen
  \bibfield  {author} {\bibinfo {author} {\bibfnamefont {S.}~\bibnamefont
  {Sarang}}, \bibinfo {author} {\bibfnamefont {O.}~\bibnamefont {Kitzler}},
  \bibinfo {author} {\bibfnamefont {O.}~\bibnamefont {Lux}}, \bibinfo {author}
  {\bibfnamefont {Z.}~\bibnamefont {Bai}}, \bibinfo {author} {\bibfnamefont
  {R.~J.}\ \bibnamefont {Williams}}, \bibinfo {author} {\bibfnamefont {D.~J.}\
  \bibnamefont {Spence}},\ and\ \bibinfo {author} {\bibfnamefont {R.~P.}\
  \bibnamefont {Mildren}},\ }\bibfield  {title} {\bibinfo {title}
  {Single-longitudinal-mode diamond laser stabilization using
  polarization-dependent raman gain},\ }\href
  {https://doi.org/10.1364/OSAC.2.001028} {\bibfield  {journal} {\bibinfo
  {journal} {OSA Continuum}\ }\textbf {\bibinfo {volume} {2}},\ \bibinfo
  {pages} {1028} (\bibinfo {year} {2019})}\BibitemShut {NoStop}%
\bibitem [{\citenamefont {Li}\ \emph {et~al.}(2020)\citenamefont {Li},
  \citenamefont {Kitzler},\ and\ \citenamefont {Spence}}]{Li:20}%
  \BibitemOpen
  \bibfield  {author} {\bibinfo {author} {\bibfnamefont {M.}~\bibnamefont
  {Li}}, \bibinfo {author} {\bibfnamefont {O.}~\bibnamefont {Kitzler}},\ and\
  \bibinfo {author} {\bibfnamefont {D.~J.}\ \bibnamefont {Spence}},\ }\bibfield
   {title} {\bibinfo {title} {Investigating single-longitudinal-mode operation
  of a continuous wave second stokes diamond raman ring laser},\ }\href
  {https://doi.org/10.1364/OE.380644} {\bibfield  {journal} {\bibinfo
  {journal} {Opt. Express}\ }\textbf {\bibinfo {volume} {28}},\ \bibinfo
  {pages} {1738} (\bibinfo {year} {2020})}\BibitemShut {NoStop}%
\bibitem [{\citenamefont {Granados}\ \emph
  {et~al.}(2022{\natexlab{a}})\citenamefont {Granados}, \citenamefont
  {Granados}, \citenamefont {Ahmed}, \citenamefont {Chrysalidis}, \citenamefont
  {Fedosseev}, \citenamefont {Marsh}, \citenamefont {Wilkins}, \citenamefont
  {Mildren},\ and\ \citenamefont {Spence}}]{Granados22}%
  \BibitemOpen
  \bibfield  {author} {\bibinfo {author} {\bibfnamefont {E.}~\bibnamefont
  {Granados}}, \bibinfo {author} {\bibfnamefont {C.}~\bibnamefont {Granados}},
  \bibinfo {author} {\bibfnamefont {R.}~\bibnamefont {Ahmed}}, \bibinfo
  {author} {\bibfnamefont {K.}~\bibnamefont {Chrysalidis}}, \bibinfo {author}
  {\bibfnamefont {V.~N.}\ \bibnamefont {Fedosseev}}, \bibinfo {author}
  {\bibfnamefont {B.~A.}\ \bibnamefont {Marsh}}, \bibinfo {author}
  {\bibfnamefont {S.~G.}\ \bibnamefont {Wilkins}}, \bibinfo {author}
  {\bibfnamefont {R.~P.}\ \bibnamefont {Mildren}},\ and\ \bibinfo {author}
  {\bibfnamefont {D.~J.}\ \bibnamefont {Spence}},\ }\bibfield  {title}
  {\bibinfo {title} {Spectral synthesis of multimode lasers to the fourier
  limit in integrated fabry--perot diamond resonators},\ }\href
  {https://doi.org/10.1364/OPTICA.447380} {\bibfield  {journal} {\bibinfo
  {journal} {Optica}\ }\textbf {\bibinfo {volume} {9}},\ \bibinfo {pages} {317}
  (\bibinfo {year} {2022}{\natexlab{a}})}\BibitemShut {NoStop}%
\bibitem [{\citenamefont {Granados}\ \emph
  {et~al.}(2022{\natexlab{b}})\citenamefont {Granados}, \citenamefont
  {Stoikos}, \citenamefont {Echarri}, \citenamefont {Chrysalidis},
  \citenamefont {Fedosseev}, \citenamefont {Granados}, \citenamefont {Leask},
  \citenamefont {Marsh},\ and\ \citenamefont
  {Mildren}}]{doi:10.1063/5.0088592}%
  \BibitemOpen
  \bibfield  {author} {\bibinfo {author} {\bibfnamefont {E.}~\bibnamefont
  {Granados}}, \bibinfo {author} {\bibfnamefont {G.}~\bibnamefont {Stoikos}},
  \bibinfo {author} {\bibfnamefont {D.~T.}\ \bibnamefont {Echarri}}, \bibinfo
  {author} {\bibfnamefont {K.}~\bibnamefont {Chrysalidis}}, \bibinfo {author}
  {\bibfnamefont {V.~N.}\ \bibnamefont {Fedosseev}}, \bibinfo {author}
  {\bibfnamefont {C.}~\bibnamefont {Granados}}, \bibinfo {author}
  {\bibfnamefont {V.}~\bibnamefont {Leask}}, \bibinfo {author} {\bibfnamefont
  {B.~A.}\ \bibnamefont {Marsh}},\ and\ \bibinfo {author} {\bibfnamefont
  {R.~P.}\ \bibnamefont {Mildren}},\ }\bibfield  {title} {\bibinfo {title}
  {Tunable spectral squeezers based on monolithically integrated diamond raman
  resonators},\ }\href {https://doi.org/10.1063/5.0088592} {\bibfield
  {journal} {\bibinfo  {journal} {Applied Physics Letters}\ }\textbf {\bibinfo
  {volume} {120}},\ \bibinfo {pages} {151101} (\bibinfo {year}
  {2022}{\natexlab{b}})},\ \Eprint
  {https://arxiv.org/abs/https://doi.org/10.1063/5.0088592}
  {https://doi.org/10.1063/5.0088592} \BibitemShut {NoStop}%
\bibitem [{\citenamefont {Spence}(2017)}]{SPENCE20171}%
  \BibitemOpen
  \bibfield  {author} {\bibinfo {author} {\bibfnamefont {D.~J.}\ \bibnamefont
  {Spence}},\ }\bibfield  {title} {\bibinfo {title} {Spectral effects of
  stimulated raman scattering in crystals},\ }\href
  {https://doi.org/https://doi.org/10.1016/j.pquantelec.2016.11.001} {\bibfield
   {journal} {\bibinfo  {journal} {Progress in Quantum Electronics}\ }\textbf
  {\bibinfo {volume} {51}},\ \bibinfo {pages} {1} (\bibinfo {year}
  {2017})}\BibitemShut {NoStop}%
\bibitem [{\citenamefont {Dzhotyan}\ \emph {et~al.}(1977)\citenamefont
  {Dzhotyan}, \citenamefont {D{\textquotesingle}yakov}, \citenamefont
  {Zubarev}, \citenamefont {Mironov},\ and\ \citenamefont
  {Mikha{\u{\i}}lov}}]{Dzhotyan_1977}%
  \BibitemOpen
  \bibfield  {author} {\bibinfo {author} {\bibfnamefont {G.~P.}\ \bibnamefont
  {Dzhotyan}}, \bibinfo {author} {\bibfnamefont {Y.~E.}\ \bibnamefont
  {D{\textquotesingle}yakov}}, \bibinfo {author} {\bibfnamefont {I.~G.}\
  \bibnamefont {Zubarev}}, \bibinfo {author} {\bibfnamefont {A.~B.}\
  \bibnamefont {Mironov}},\ and\ \bibinfo {author} {\bibfnamefont {S.~I.}\
  \bibnamefont {Mikha{\u{\i}}lov}},\ }\bibfield  {title} {\bibinfo {title}
  {Influence of the spectral width and statistics of a stokes signal on the
  efficiency of stimulated raman scattering of nonmonochromatic pump
  radiation},\ }\href {https://doi.org/10.1070/qe1977v007n06abeh012901}
  {\bibfield  {journal} {\bibinfo  {journal} {Soviet Journal of Quantum
  Electronics}\ }\textbf {\bibinfo {volume} {7}},\ \bibinfo {pages} {783}
  (\bibinfo {year} {1977})}\BibitemShut {NoStop}%
\bibitem [{\citenamefont {Sidorovich}(1978)}]{Sidorovich_1978}%
  \BibitemOpen
  \bibfield  {author} {\bibinfo {author} {\bibfnamefont {V.~G.}\ \bibnamefont
  {Sidorovich}},\ }\bibfield  {title} {\bibinfo {title} {Reproduction of the
  pump spectrum in stimulated raman scattering},\ }\href
  {https://doi.org/10.1070/qe1978v008n06abeh010403} {\bibfield  {journal}
  {\bibinfo  {journal} {Soviet Journal of Quantum Electronics}\ }\textbf
  {\bibinfo {volume} {8}},\ \bibinfo {pages} {784} (\bibinfo {year}
  {1978})}\BibitemShut {NoStop}%
\bibitem [{\citenamefont {Trutna}\ \emph {et~al.}(1979)\citenamefont {Trutna},
  \citenamefont {Park},\ and\ \citenamefont {Byer}}]{1070054}%
  \BibitemOpen
  \bibfield  {author} {\bibinfo {author} {\bibfnamefont {W.}~\bibnamefont
  {Trutna}}, \bibinfo {author} {\bibfnamefont {Y.}~\bibnamefont {Park}},\ and\
  \bibinfo {author} {\bibfnamefont {R.}~\bibnamefont {Byer}},\ }\bibfield
  {title} {\bibinfo {title} {The dependence of raman gain on pump laser
  bandwidth},\ }\href {https://doi.org/10.1109/JQE.1979.1070054} {\bibfield
  {journal} {\bibinfo  {journal} {IEEE Journal of Quantum Electronics}\
  }\textbf {\bibinfo {volume} {15}},\ \bibinfo {pages} {648} (\bibinfo {year}
  {1979})}\BibitemShut {NoStop}%
\bibitem [{\citenamefont {Warner}\ and\ \citenamefont
  {Bobbs}(1986)}]{Warner:86}%
  \BibitemOpen
  \bibfield  {author} {\bibinfo {author} {\bibfnamefont {C.}~\bibnamefont
  {Warner}}\ and\ \bibinfo {author} {\bibfnamefont {B.}~\bibnamefont {Bobbs}},\
  }\bibfield  {title} {\bibinfo {title} {Effects of off-resonant raman
  interactions on multimode stokes conversion efficiency and output wave
  front},\ }\href {https://doi.org/10.1364/JOSAB.3.001345} {\bibfield
  {journal} {\bibinfo  {journal} {J. Opt. Soc. Am. B}\ }\textbf {\bibinfo
  {volume} {3}},\ \bibinfo {pages} {1345} (\bibinfo {year} {1986})}\BibitemShut
  {NoStop}%
\bibitem [{\citenamefont {Westling}\ and\ \citenamefont
  {Raymer}(1987)}]{PhysRevA.36.4835}%
  \BibitemOpen
  \bibfield  {author} {\bibinfo {author} {\bibfnamefont {L.~A.}\ \bibnamefont
  {Westling}}\ and\ \bibinfo {author} {\bibfnamefont {M.~G.}\ \bibnamefont
  {Raymer}},\ }\bibfield  {title} {\bibinfo {title} {Intensity correlation
  measurements in stimulated raman generation with a multimode laser},\ }\href
  {https://doi.org/10.1103/PhysRevA.36.4835} {\bibfield  {journal} {\bibinfo
  {journal} {Phys. Rev. A}\ }\textbf {\bibinfo {volume} {36}},\ \bibinfo
  {pages} {4835} (\bibinfo {year} {1987})}\BibitemShut {NoStop}%
\bibitem [{\citenamefont {Xiong}\ \emph {et~al.}(2007)\citenamefont {Xiong},
  \citenamefont {Murphy}, \citenamefont {Carlsten},\ and\ \citenamefont
  {Repasky}}]{Xiong:07}%
  \BibitemOpen
  \bibfield  {author} {\bibinfo {author} {\bibfnamefont {Y.}~\bibnamefont
  {Xiong}}, \bibinfo {author} {\bibfnamefont {S.}~\bibnamefont {Murphy}},
  \bibinfo {author} {\bibfnamefont {J.~L.}\ \bibnamefont {Carlsten}},\ and\
  \bibinfo {author} {\bibfnamefont {K.}~\bibnamefont {Repasky}},\ }\bibfield
  {title} {\bibinfo {title} {Theory of a far-off resonance mode-locked raman
  laser in h2 with high finesse cavity enhancement},\ }\href
  {https://doi.org/10.1364/JOSAB.24.002055} {\bibfield  {journal} {\bibinfo
  {journal} {J. Opt. Soc. Am. B}\ }\textbf {\bibinfo {volume} {24}},\ \bibinfo
  {pages} {2055} (\bibinfo {year} {2007})}\BibitemShut {NoStop}%
\bibitem [{\citenamefont {Jacobson}\ and\ \citenamefont
  {Stoupin}(2019)}]{JACOBSON2019107469}%
  \BibitemOpen
  \bibfield  {author} {\bibinfo {author} {\bibfnamefont {P.}~\bibnamefont
  {Jacobson}}\ and\ \bibinfo {author} {\bibfnamefont {S.}~\bibnamefont
  {Stoupin}},\ }\bibfield  {title} {\bibinfo {title} {Thermal expansion
  coefficient of diamond in a wide temperature range},\ }\href
  {https://doi.org/https://doi.org/10.1016/j.diamond.2019.107469} {\bibfield
  {journal} {\bibinfo  {journal} {Diamond and Related Materials}\ }\textbf
  {\bibinfo {volume} {97}},\ \bibinfo {pages} {107469} (\bibinfo {year}
  {2019})}\BibitemShut {NoStop}%
\bibitem [{\citenamefont {Hervé}\ and\ \citenamefont
  {Vandamme}(1994)}]{HERVE1994609}%
  \BibitemOpen
  \bibfield  {author} {\bibinfo {author} {\bibfnamefont {P.}~\bibnamefont
  {Hervé}}\ and\ \bibinfo {author} {\bibfnamefont {L.}~\bibnamefont
  {Vandamme}},\ }\bibfield  {title} {\bibinfo {title} {General relation between
  refractive index and energy gap in semiconductors},\ }\href
  {https://doi.org/https://doi.org/10.1016/1350-4495(94)90026-4} {\bibfield
  {journal} {\bibinfo  {journal} {Infrared Physics \& Technology}\ }\textbf
  {\bibinfo {volume} {35}},\ \bibinfo {pages} {609} (\bibinfo {year}
  {1994})}\BibitemShut {NoStop}%
\bibitem [{\citenamefont {Loudon}(1964)}]{doi:10.1080/00018736400101051}%
  \BibitemOpen
  \bibfield  {author} {\bibinfo {author} {\bibfnamefont {R.}~\bibnamefont
  {Loudon}},\ }\bibfield  {title} {\bibinfo {title} {The raman effect in
  crystals},\ }\href {https://doi.org/10.1080/00018736400101051} {\bibfield
  {journal} {\bibinfo  {journal} {Advances in Physics}\ }\textbf {\bibinfo
  {volume} {13}},\ \bibinfo {pages} {423} (\bibinfo {year} {1964})},\ \Eprint
  {https://arxiv.org/abs/https://doi.org/10.1080/00018736400101051}
  {https://doi.org/10.1080/00018736400101051} \BibitemShut {NoStop}%
\bibitem [{\citenamefont {Gonzalez}\ \emph {et~al.}(1996)\citenamefont
  {Gonzalez}, \citenamefont {Moya},\ and\ \citenamefont {Chervin}}]{article}%
  \BibitemOpen
  \bibfield  {author} {\bibinfo {author} {\bibfnamefont {J.}~\bibnamefont
  {Gonzalez}}, \bibinfo {author} {\bibfnamefont {E.}~\bibnamefont {Moya}},\
  and\ \bibinfo {author} {\bibfnamefont {J.}~\bibnamefont {Chervin}},\
  }\bibfield  {title} {\bibinfo {title} {Anharmonic effects in light scattering
  due to optical phonons in cugas\_{2}},\ }\href
  {https://doi.org/10.1103/PhysRevB.54.4707} {\bibfield  {journal} {\bibinfo
  {journal} {Phys. Rev. B}\ }\textbf {\bibinfo {volume} {54}} (\bibinfo {year}
  {1996})}\BibitemShut {NoStop}%
\bibitem [{\citenamefont {Turri}\ \emph {et~al.}(2017)\citenamefont {Turri},
  \citenamefont {Webster}, \citenamefont {Chen}, \citenamefont {Wickham},
  \citenamefont {Bennett},\ and\ \citenamefont {Bass}}]{Turri:17}%
  \BibitemOpen
  \bibfield  {author} {\bibinfo {author} {\bibfnamefont {G.}~\bibnamefont
  {Turri}}, \bibinfo {author} {\bibfnamefont {S.}~\bibnamefont {Webster}},
  \bibinfo {author} {\bibfnamefont {Y.}~\bibnamefont {Chen}}, \bibinfo {author}
  {\bibfnamefont {B.}~\bibnamefont {Wickham}}, \bibinfo {author} {\bibfnamefont
  {A.}~\bibnamefont {Bennett}},\ and\ \bibinfo {author} {\bibfnamefont
  {M.}~\bibnamefont {Bass}},\ }\bibfield  {title} {\bibinfo {title} {Index of
  refraction from the near-ultraviolet to the near-infrared from a single
  crystal microwave-assisted cvd diamond},\ }\href
  {https://doi.org/10.1364/OME.7.000855} {\bibfield  {journal} {\bibinfo
  {journal} {Opt. Mater. Express}\ }\textbf {\bibinfo {volume} {7}},\ \bibinfo
  {pages} {855} (\bibinfo {year} {2017})}\BibitemShut {NoStop}%
\bibitem [{\citenamefont {Klemens}(1966)}]{PhysRev.148.845}%
  \BibitemOpen
  \bibfield  {author} {\bibinfo {author} {\bibfnamefont {P.~G.}\ \bibnamefont
  {Klemens}},\ }\bibfield  {title} {\bibinfo {title} {Anharmonic decay of
  optical phonons},\ }\href {https://doi.org/10.1103/PhysRev.148.845}
  {\bibfield  {journal} {\bibinfo  {journal} {Phys. Rev.}\ }\textbf {\bibinfo
  {volume} {148}},\ \bibinfo {pages} {845} (\bibinfo {year}
  {1966})}\BibitemShut {NoStop}%
\bibitem [{\citenamefont {Debernardi}\ \emph {et~al.}(1995)\citenamefont
  {Debernardi}, \citenamefont {Baroni},\ and\ \citenamefont
  {Molinari}}]{PhysRevLett.75.1819}%
  \BibitemOpen
  \bibfield  {author} {\bibinfo {author} {\bibfnamefont {A.}~\bibnamefont
  {Debernardi}}, \bibinfo {author} {\bibfnamefont {S.}~\bibnamefont {Baroni}},\
  and\ \bibinfo {author} {\bibfnamefont {E.}~\bibnamefont {Molinari}},\
  }\bibfield  {title} {\bibinfo {title} {Anharmonic phonon lifetimes in
  semiconductors from density-functional perturbation theory},\ }\href
  {https://doi.org/10.1103/PhysRevLett.75.1819} {\bibfield  {journal} {\bibinfo
   {journal} {Phys. Rev. Lett.}\ }\textbf {\bibinfo {volume} {75}},\ \bibinfo
  {pages} {1819} (\bibinfo {year} {1995})}\BibitemShut {NoStop}%
\bibitem [{\citenamefont {Liu}\ \emph {et~al.}(2000)\citenamefont {Liu},
  \citenamefont {Bursill}, \citenamefont {Prawer},\ and\ \citenamefont
  {Beserman}}]{PhysRevB.61.3391}%
  \BibitemOpen
  \bibfield  {author} {\bibinfo {author} {\bibfnamefont {M.~S.}\ \bibnamefont
  {Liu}}, \bibinfo {author} {\bibfnamefont {L.~A.}\ \bibnamefont {Bursill}},
  \bibinfo {author} {\bibfnamefont {S.}~\bibnamefont {Prawer}},\ and\ \bibinfo
  {author} {\bibfnamefont {R.}~\bibnamefont {Beserman}},\ }\bibfield  {title}
  {\bibinfo {title} {Temperature dependence of the first-order raman phonon
  line of diamond},\ }\href {https://doi.org/10.1103/PhysRevB.61.3391}
  {\bibfield  {journal} {\bibinfo  {journal} {Phys. Rev. B}\ }\textbf {\bibinfo
  {volume} {61}},\ \bibinfo {pages} {3391} (\bibinfo {year}
  {2000})}\BibitemShut {NoStop}%
\bibitem [{\citenamefont {Leask}(2019)}]{thesis}%
  \BibitemOpen
  \bibfield  {author} {\bibinfo {author} {\bibfnamefont {V.}~\bibnamefont
  {Leask}},\ }\emph {\bibinfo {title} {A Continuously Tunable Single
  Longitudinal Mode Diamond Raman Laser}},\ \href@noop {} {Master's thesis},\
  \bibinfo  {school} {SUPA Department of Physics, University of Strathclyde},
  \bibinfo {address} {Glasgow G4 0NG} (\bibinfo {year} {2019})\BibitemShut
  {NoStop}%
\bibitem [{\citenamefont {Moelle}\ \emph {et~al.}(1997)\citenamefont {Moelle},
  \citenamefont {Klose}, \citenamefont {Szücs}, \citenamefont {Fecht},
  \citenamefont {Johnston}, \citenamefont {Chalker},\ and\ \citenamefont
  {Werner}}]{MOELLE1997839}%
  \BibitemOpen
  \bibfield  {author} {\bibinfo {author} {\bibfnamefont {C.}~\bibnamefont
  {Moelle}}, \bibinfo {author} {\bibfnamefont {S.}~\bibnamefont {Klose}},
  \bibinfo {author} {\bibfnamefont {F.}~\bibnamefont {Szücs}}, \bibinfo
  {author} {\bibfnamefont {H.}~\bibnamefont {Fecht}}, \bibinfo {author}
  {\bibfnamefont {C.}~\bibnamefont {Johnston}}, \bibinfo {author}
  {\bibfnamefont {P.}~\bibnamefont {Chalker}},\ and\ \bibinfo {author}
  {\bibfnamefont {M.}~\bibnamefont {Werner}},\ }\bibfield  {title} {\bibinfo
  {title} {Measurement and calculation of the thermal expansion coefficient of
  diamond},\ }\href
  {https://doi.org/https://doi.org/10.1016/S0925-9635(96)00674-7} {\bibfield
  {journal} {\bibinfo  {journal} {Diamond and Related Materials}\ }\textbf
  {\bibinfo {volume} {6}},\ \bibinfo {pages} {839} (\bibinfo {year}
  {1997})}\BibitemShut {NoStop}%
\bibitem [{ind()}]{index}%
  \BibitemOpen
  \bibfield  {title} {\bibinfo {title} {{Diamond Materials, THE CVD Diamond
  Booklet}},\ }\href@noop {} {\bibinfo  {journal} {{Freiburg: Diamond
  Materials, Advanced Diamond Technology}}\ }\BibitemShut {NoStop}%
\end{thebibliography}%

\end{document}